\documentclass{aa}  

\usepackage{graphicx}
\usepackage{txfonts}
\usepackage{lipsum}
\usepackage{booktabs}
\usepackage{siunitx}
\usepackage{subcaption}         
\usepackage{lscape}             
\usepackage{placeins}           
                                
\begin{document}

   \title{Coupling dynamical accretion and chemical differentiation:\\ A unified framework for the diversity of Earth and Mars}

   \author{Zhihui Kong\inst{1,2}
        \and Beibei Liu\inst{1,2}\thanks{Corresponding authors: Beibei Liu (bbliu@zju.edu.cn)}
        \and J. ZhangZhou\inst{3}
        \and Haolan Tang\inst{4}
        \and Zhengbin Deng\inst{4}
        \and Qun-Ke Xia \inst{3}
        \and Simon L. Grimm \inst{5,6}
        \and Man Hoi Lee \inst{7,8,9}
        \and Yi Huang \inst{1,2}
        \and Liping Qin \inst{3}
        \and Jonathan H. Jiang \inst{10}
        }

   \institute{Institute for Astronomy, School of Physics, Zhejiang University, Hangzhou 310027, China e-mail:bbliu@zju.edu.cn
   \and Center for Cosmology and Computational Astrophysics, Institute for Advanced Study in Physics, Zhejiang University, Hangzhou 310027, China
   \and Research Center for Earth and Planetary Material Sciences, School of Earth Sciences, Zhejiang University, Hangzhou, 310058, China
   \and State Key Laboratory of Lithospheric and Environmental Coevolution, University of Science and Technology of China, Hefei, 230026, China
   \and Institute for Particle Physics and Astrophysics, ETH Zurich, CH-8093 Zurich, Switzerland
   \and Department of Astrophysics, University of Zurich, CH-8057 Zurich, Switzerland
   \and Department of Earth and Planetary Sciences, The University of Hong Kong, Pokfulam Road, Hong Kong, China
   \and Department of Physics, The University of Hong Kong, Pokfulam Road, Hong Kong, China
   \and Hong Kong Institute for Astronomy and Astrophysics, Hong Kong, China
   \and California Institute of Technology, Pasadena, California 91106, U.S.A.
   }

   \date{TBD}

  \abstract
    {The physical and geochemical differences between Earth and Mars provide fundamental constraints on terrestrial planet formation, yet a self-consistent framework linking dynamical and chemical aspects remains elusive. Here we present an integrated modeling framework that couples high-resolution N-body simulations with impact-driven metal-silicate equilibration to track the dynamical accretion history and chemical differentiation for Earth and Mars.
    Using a narrow ring planetesimal accretion scenario, we show that Earth and Mars analogs naturally sample systematically different solid reservoirs within the protoplanetary disk. Earth analogs preferentially accrete reduced material around the planetesimal ring center, whereas Mars analogs acquire a larger fraction of oxidized material exterior to the ring.  This leads to diverse bulk redox states, with composition further modified by impact-dependent pressure-temperature equilibration conditions during core formation. As a result, Earth analogs experience deeper equilibration and more efficient transfer of iron into the core, producing mantles with low iron oxide contents and  larger core mass fractions. In contrast, Mars analogs equilibrate at shallower conditions, retain more iron in their mantles, and develop smaller cores.
    Our results demonstrate that the dynamical and geochemical differences between Earth and Mars emerge from the coupled effects of accretion pathways, the disk's radial redox structure, and impact-controlled differentiation rather than from any single process. Our unified framework physically explains the geochemical diversity of terrestrial planets and offers a potential pathway to interpret compositions of rocky planets in exoplanetary systems.
}


\authorrunning{Kong et al.}
\titlerunning{Coupled accretion-differentiation: Earth and Mars}
\maketitle
\nolinenumbers

\section{Introduction}

The distinct physical and geochemical properties of Earth and Mars stand as one of the central questions in astrophysics and planetary science. These two neighboring terrestrial planets, although originating from a common region in the early Solar System, display clear differences.
A striking feature of the Earth-Mars contrast is their large mass ratio, $M_{\rm Earth}/M_{\rm Mars}{\approx}10$.
Furthermore, Earth's bulk silicate is  more reduced, with ${\sim}8\%$ FeO, whereas Mars has a globally oxidized mantle, with ${\sim}14\%$ FeO. Earth's core contains moderate amounts of light elements (Si, O) and has a core mass fraction of ${\sim}32\%$, whereas Mars has a smaller core mass fraction of $18{-}25\%$ \citep{wanke_1994,morgan_1979,wade_2005,taylor_2013,yoshizaki_2020,Yoshizaki_2021,Guimond_2023}. Radiometric dating further indicates that Mars formed early, likely within $2{-}4$ Myr after Calcium-Aluminum-rich Inclusions (CAIs) \citep{Dauphas2011,Tang_2014}, whereas Earth accumulated more slowly (within $30{-}150$ Myr) and experienced prolonged differentiation \citep{touboul_2007,kleine_2011}. 

Astrophysical N-body simulations have advanced our understanding of terrestrial planet formation by exploring the influences of feeding-zone distributions, radial-mixing efficiency, and giant-planet architecture \citep{raymond_2004a,hansen_2009a,nesvorny_2021b,woo_2022a,batygin_2023b,kong_2024,clement_2024}. However, the studies employing these simulations typically do not explicitly resolve the evolution of planetary bulk redox state and/or chemical element partitioning during accretion. Conversely, geochemical models reconstructing core-mantle differentiation often rely on prescribed oxidation gradients or idealized sequences of impact-driven equilibration \citep{wade_2005,rubie_2011,siebert_2013,Dauphas_2017,brennan_2020} rather than integrating realistic accretion histories from dynamical simulations.

In summary, dynamical evolution and geochemical fractionation are often treated as isolated processes, preventing the simultaneous reproduction of planetary growth and chemical element abundance in a self-consistent framework. While several studies have developed dynamical-geochemical coupled models to investigate (primarily) Earth's accretion history \citep{rubie_2015,fischer_2015,Rubie_2016,fischer_2017,Jennings_2021,Blanchard_2022,dale_2023,gu_2023,dale_2025,rubie_2025}, with less attention given to Mars \citep{brennan_2020,Nathan_2023}, none have attempted to simultaneously reproduce and compare the chemical properties of both Earth and Mars within a unified dynamical framework. Consequently, the origin of the Earth-Mars geochemical divergence remains poorly constrained, especially because most models treat accretion dynamics and chemical differentiation separately or focus on individual planets.

Motivated by this gap in research, we present an integrated model that couples N-body dynamics, impact-driven melting, element partitioning, redox evolution, and multistage core-mantle differentiation in a single unified framework. Our goals are twofold: (1) to identify accretion pathways that successfully reproduce Earth and Mars analogs, and (2) to quantify the controlling mechanisms underlying their key geochemical properties. By unifying these interconnected processes, this approach enables quantitative tracking of chemistry-dynamics coevolution during terrestrial planet formation, allowing valuable constraints on the distinctive features of Earth and Mars to be obtained.

The paper is organized as follows. Section 2 describes the integrated modeling framework. Section 3 presents key results for our simulated Earth and Mars analogs, including direct comparisons to observational geochemical constraints. In Section 4 we discuss critical model assumptions and the physical mechanisms driving the distinct geochemical features of Earth and Mars. Finally, we summarize the key findings in Section 5.  

\section{Methods}
\subsection{Narrow ring model of terrestrial planet formation}
\subsubsection{Planetesimal disk profile}
For this study, we adopted the narrow-ring model of terrestrial planet formation. This configuration has been shown to efficiently produce Earth-sized planets near the ring center while simultaneously forming Mars-sized bodies close to its outer edge \citep{hansen_2009a,nesvorny_2021b,izidoro_2022a,Woo_2023,woo_2024}. 

\begin{figure}
\centering
\includegraphics[width=0.95\linewidth]{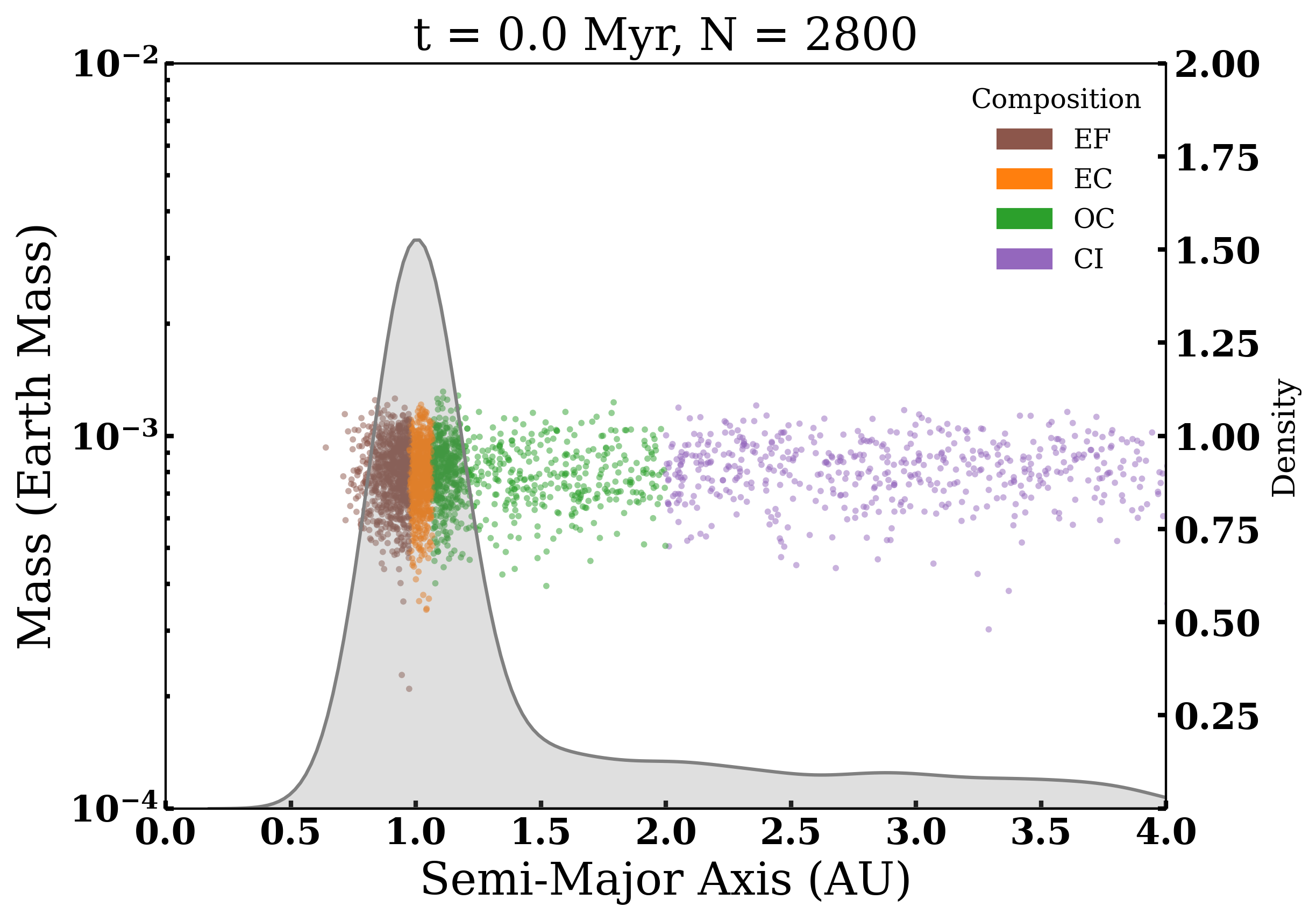}
\caption{Initial radial distribution of compositional endmembers at  t = 0.0 Myr in our simulations corresponds to the early epoch of planetesimal differentiation, which occurred approximately $\sim$1--2 Myr after the formation of CAIs. EF and EC materials are separated at the left-side two-fifths division point of the Gaussian annulus (a=0.9747 AU), while the EC and OC boundary is defined at the right-side one-fourth division point (a=1.0674 AU). The OC/CI transition is located at 2.0 AU. Colored points denote different compositional tags, and the gray curve (right axis) shows the normalized number-density distribution of planetesimals.}\label{Fig1}
\end{figure}

We initialized 2000 planetesimals in a narrow annulus centered at 1 AU. Their radial distances were drawn from a Gaussian distribution:
\begin{equation}
p(a) = \frac{1}{\sqrt{2\pi} \sigma}
\exp\left[-\frac{(a - 1 \ \mathrm{AU})^{2}}{2\sigma^{2}}\right],
\qquad
\sigma = 0.1 \ \mathrm{AU}.
\end{equation}
An additional population of 800 planetesimals was distributed between 1.3 and 4.0 AU following an $r^{-1}$ surface density profile. All planetesimals were treated as active particles, with full gravitational interactions included. The masses, whether within the ring or the disk, were sampled from a Gaussian distribution characterized by a mean value of $8\times 10^{-4} \,\rm M_{\oplus}$ and a standard deviation of $1.6 \times 10^{-4} \,\rm M_{\oplus}$. The total masses of the planetesimals are approximately $1.6 \,\rm M_{\oplus}$ within the ring and approximately $0.64 \,\rm M_{\oplus}$ within the disk.
The initial orbital eccentricities and inclinations were from Rayleigh distributions with scale values ($e_0$ and $i_0$) of 0.005 and 0.0025.  Fig.~\ref{Fig1} illustrates the initial mass and radial distribution of planetesimals in our model, with color representing the different compositional groups (such as enstatite chondrites (EC), ordinary chondrites (OC), Ivuna-type carbonaceous chondrites (CI), and a hypothetical composition termed EF; see details in Section 2.2).

Meanwhile, we placed Jupiter and Saturn near a $2$:$1$ mean motion resonance beyond the outer edge of the planetesimal disk to account for their gravitational perturbations. In our setup, the giant planets were initialized on slightly noncircular and inclined orbits: Jupiter at 5.45 AU, with $e {=} 0.009$ and $i {=} 0^\circ$, and Saturn at 8.81 AU, with $e {= }0.0002$ and $i {=} 0.5^\circ$. This configuration is consistent with those used in early Solar System evolution models \citep{tsiganis_2005,Woo_2023,kong_2024}.

\subsubsection{Gas disk profile}
Mars is inferred to have formed very rapidly, within only a few million years \citep{touboul_2007, nimmo_2007, kleine_2009, dauphas_2011, Tang_2014}, and is widely regarded as a stranded planetary embryo that did not undergo substantial late-stage accretion. This rapid accretion timescale coincides with the dissipation of the protoplanetary gas disk \citep{haisch_2001, Mamajek_2009}, implying that gas dispersal played a critical role in shaping the dynamical environment of the inner Solar System. Consequently, modeling the simultaneous formation of Earth and Mars requires explicit consideration of gas-related effects.

Here we adopted a modified minimum-mass solar nebula (MMSN) disk profile \citep{weidenschilling_1977, hayashi_1981a, armitage_2011a} by incorporating the effects of magnetically driven winds \citep{Suzuki_2016,Ogihara_2018a}. The resulting gas disk structure is parameterized as  
\begin{equation}
\begin{aligned}
\Sigma_{\mathrm{gas}}(r,t) &= \Sigma_{\mathrm{gas},0}
\left(\frac{r}{1\,\mathrm{AU}}\right)^{\beta(r)}
\exp\!\left(-\frac{t}{\tau_{\mathrm{decay}}}\right), \\
\beta(r) &= - \ln\left(\frac{r}{1\,\mathrm{AU}}\right) - 0.5,
\end{aligned}
\end{equation}
where $r$ is the heliocentric distance from the Sun, and we adopted $\tau_{\mathrm{decay}} {=} 1$ Myr for the gas disk dissipation timescale (all remaining disk gas was removed at $t {=} 5$ Myr) and $\Sigma_{\mathrm{gas},0}{=}1700 \ \rm g\ cm^{-2}$. This prescription accounts for the effects of magnetically driven winds, which preferentially deplete the inner disk and produce a radially structured gas surface density profile \citep{Nesvorny_2025}. The time zero of our simulations corresponds to the early planetesimal differentiation epoch, which occurred within $\sim$1--2 Myr after CAI formation. 

The disk gas influences the dynamics of planets and planetesimals through aerodynamic drag, orbital damping, and type-I migration \citep{weidenschilling_1977,Morishima_2010}. Planetesimals experience aerodynamic gas drag, while protoplanets undergo type-I orbital migration as well as eccentricity and inclination damping. All bodies were initially assumed to have a constant bulk density of $3~\mathrm{g\,cm^{-3}}$. In addition, we included the gravitational potential of the gas disk, which modifies the apsidal precession of planetary bodies during gas disk dispersal (e.g.,  \citealt{Shuai2025}). The prescriptions adopted for these gas-related effects follow \cite{Woo_2021}.

\subsection{Chemical composition and radial distribution of planetesimals}
In this study, we focus on a set of major elements, specifically Mg, Al, Ca, Fe, Ni, Si, and O. These elements account for most of the mass of terrestrial planets, and their partitioning behavior between metallic and silicate phases (a critical process in planetary differentiation) is well constrained by experimental and observational studies. Volatile elements are excluded from the present analysis, as our primary goal is to establish a first-order geochemical framework. Future work will expand this analysis to incorporate volatile species and trace elements such as C, H and S.

To chemically tag particles in our N-body simulations, we adopted bulk compositions of key meteoritic parent bodies: EC, OC and CI. In addition, we incorporated a hypothetical composition for the innermost disk region, termed EF (see \citealt{dale_2025}). This EF component is modeled as refractory-rich, with enhanced Al and Ca but depleted in Si, reflecting the highly reduced, high-temperature conditions expected at the smallest heliocentric distances. The inclusion of this composition is motivated by the lower condensation and melting temperatures of Si relative to more refractory lithophile elements. The current absence of meteorites representing this group may result from its efficient accretion during terrestrial planet formation. The complete set of adopted elemental abundances is provided in Appendix A. Note that the compositions listed in Table~\ref{TabS1} are expressed as normalized compositional reservoirs rather than self-consistent planetary bodies. They define elemental ratios available for accretional mixing and subsequent differentiation and therefore do not correspond to unique core mass fractions prior to the modeled metal-silicate equilibration. 

In our model, Mg, Al, and Ca are treated as highly lithophile elements and assigned entirely in the silicate mantle. In comparison, Fe and Ni are predominantly siderophile, partitioning strongly into the metallic core. The behavior of Si and O, however, is redox-sensitive; these elements are moderately siderophile elements, with their distribution between metal and silicate governed by the system's oxygen fugacity. The initial redox states of the four groups of planetesimals  follow Table 3 of \cite{dale_2025}. We further assumed that CI planetesimals represent primitive, undifferentiated material, whereas the EF, EC, and OC groups have experienced prior metal-silicate segregation consistent with their redox states.\footnote{For initial metal-silicate equilibration, we adopt a pressure of 0.65 times the core-mantle boundary pressure from Table 2 of \cite{dale_2025}. Note that the resultant terrestrial planet compositions are influenced far more strongly by the initial redox state than by the assumed $P$--$T$ conditions.} This assumption is motivated by the near-solar composition and lack of differentiation signatures in CI chondrites, which are widely interpreted as chemically primitive Solar System material.

\begin{figure}
\centering
\sidecaption
\includegraphics[width=\linewidth]{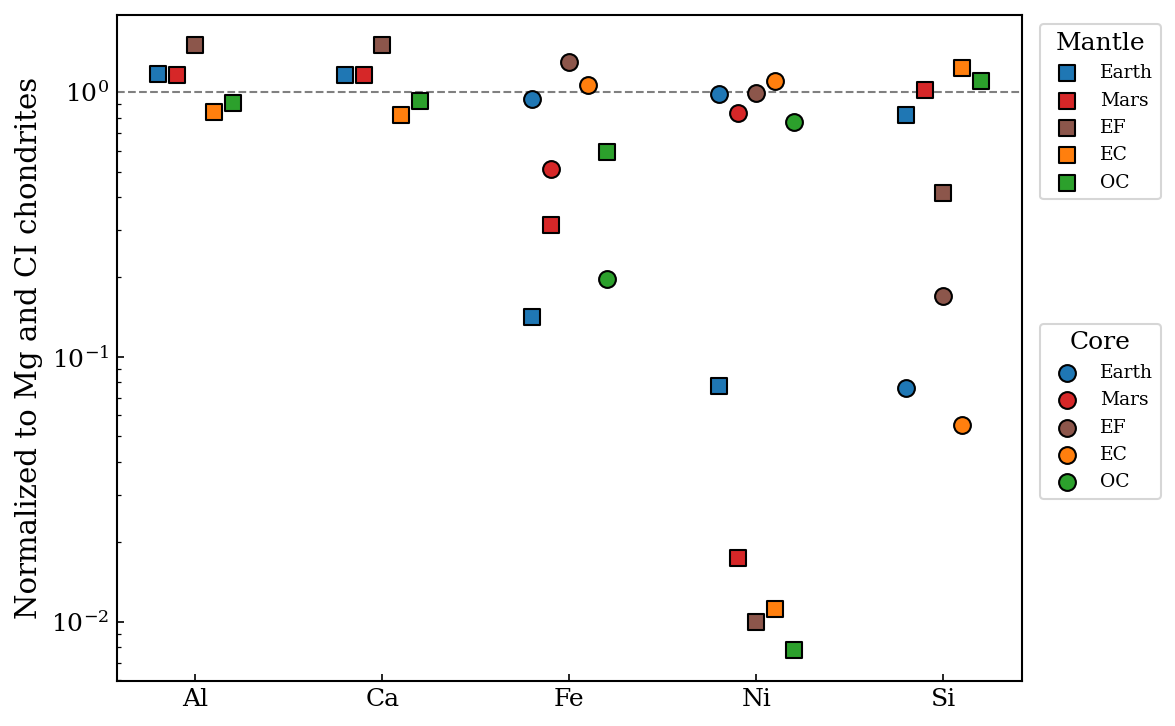}
\caption{Elemental abundances of Al, Ca, Fe, Ni, and Si, expressed in atomic percent and normalized to Mg and CI chondrites. Squares indicate mantle compositions and circles indicate core compositions, respectively. Blue and red symbols correspond to geochemically inferred present-day compositions Earth and Mars (including the Bulk Silicate Earth, Earth's core, the Bulk Silicate Mars, and the Martian core), which serve as observational reference values. In contrast, the meteoritic end-members (EF, EC, and OC) represent the initial compositional reservoirs adopted in the model. As Mg is used as the normalization reference, its abundance is not shown explicitly.}\label{Fig2}
\end{figure}

Figure~\ref{Fig2} illustrates the elemental distributions of these compositions, with the abundances of Al, Ca, Fe, Ni, and Si normalized to Mg and CI chondrites values. In contrast, the meteoritic end-members (EF, EC, and OC) represent the initial compositional reservoirs adopted in the model. As Mg serves as the normalization reference, its abundance is not explicitly shown. All elemental abundances are reported in atomic percent throughout this work to ensure consistency with the metal-silicate mass balance calculations described in Section 2.3.

Here the compositional end-members (EF, EC, OC, and CI) are assigned according to heliocentric distance. Within the narrow annulus centered at 1 AU, the boundary between EF and EC materials is defined at the left-side two-fifths division point of the Gaussian distribution (a=0.9747 AU), while the boundary between EC and OC materials is located at the right-side one-fourth division point (a=1.0674 AU). Planetesimals interior to 0.9747 AU are assigned the EF composition, whereas bodies between 0.9747 and 1.0674 AU are treated as EC material. Beyond the annulus, planetesimals between 1.0674 and 2.0 AU are assigned OC compositions, while those located beyond 2.0 AU are assigned CI compositions. The initial mass ratios of EF, EC, OC and CI are approximately (1:0.87:1:0.6). This systematic progression toward more oxidized and volatile-rich material with increasing heliocentric distance reflects the expected radial redox gradient of the protoplanetary disk and the transition across the snow line \citep{hayashi_1981a, warren_2011a, Oka_2011}.

For simplicity, we assume perfect mergers during collisions. The elemental abundances of colliding particles are combined via mass-weighted averaging, which allows the bulk compositions of growing planets to evolve self-consistently from their accretion histories. These compositions then serve as the initial conditions for the subsequent modeling of core-mantle differentiation in Section 2.3.

\subsection{Core-mantle differentiation}

The metal-silicate equilibration follows \cite{rubie_2015}, which self-consistently tracks the chemical differentiation along the impact-driven accretion history of the growing planets. For each equilibration, the element partitioning between metallic and silicate liquids is evaluated as a function of equilibration pressure and temperature ($P_{\rm eq}$, $T_{\rm eq}$), as well as oxygen fugacity ($f_{O_2}$).

The partition coefficient for element $i$ is written as
\begin{equation}
D_i^{\mathrm{met/sil}}(P_{\rm eq}, T_{\rm eq}, f_{O_2}) =\frac{C_i^{\mathrm{met}}}{C_i^{\mathrm{sil}}},
\end{equation}
where $C_i^{\rm met}$ and $C_i^{\rm sil}$ are the concentrations of element $i$ in the metal and silicate phases, respectively. The $P$--$T$ conditions for evaluating the partition coefficients are assigned according to the impact regimes described in Section~2.4.

Following \cite{rubie_2015}, element partitioning is parameterized using distribution coefficients normalized to Fe,
\begin{equation}
K_D^{i} = \frac{D_i^{\mathrm{met/sil}}}{\left(D_{\mathrm{Fe}}^{\mathrm{met/sil}}\right)^{n/2}},
\end{equation}
where $n$ is the valence of element $i$. The pressure and temperature dependence of $K_D^i$ is given by 
\begin{equation}
\log_{10} K_D^{i} = a + \frac{b}{T} + \frac{cP}{T},
\end{equation}
where the element-specific coefficients $a$, $b$, $c$ are adopted from the experimental calibrations summarized in Supplementary Table~S1 of \cite{fischer_2017}.

For each equilibration event, the compositions of metallic and silicate liquids are obtained by solving a coupled set of mass-balance and partitioning equations. Mass conservation requires $M_i^{\rm tot} {=} M_i^{\rm met} + M_i^{\rm sil}$ for each element $i$. 
Fe, Ni, Si, and O are conserved in the equilibrating metal-silicate system, whereas Mg, Al, and Ca remain entirely in the silicate phase. Oxygen fugacity is calculated self-consistently from O partitioning. Further details can be found in \cite{rubie_2015}.

\subsection{Impact regimes}

To determine the relevant $P$--$T$ conditions for metal-silicate equilibration, we distinguish between early- and late-stage impacts based on the effectiveness of heating by short-lived radionuclides. For simplicity, all impacts are assumed to occur at a fixed angle of $45^{\circ}$, corresponding to the mean impact geometry expected for random collisions \citep{gu_2023}. 
\begin{figure*}
\centering
\sidecaption
\includegraphics[width=0.85\linewidth]{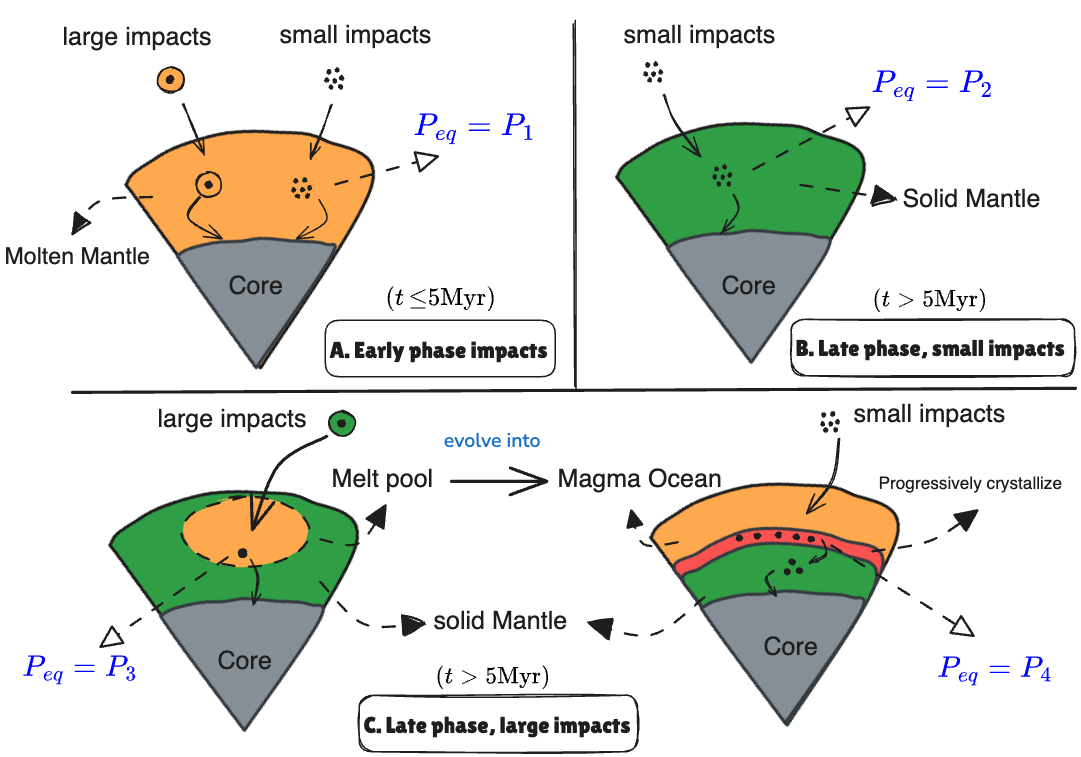}
\caption{Schematic summary of impact-dependent metal-silicate equilibration regimes during planetary accretion. (A) Early phase impacts: During the gas disk phase ($t \le 5$ Myr), embryos are assumed to be fully molten, and metal-silicate equilibration occurs within a global magma ocean at pressure $P_{1}$. (B) Late phase, small impacts: After gas disk dispersal ($t > 5$ Myr), small impacts (planetesimals) do not generate large-scale melting; equilibration occurs during metal descent through the mantle at pressure $P_{2}$. (C) Late phase, large impacts: Giant impacts (embryos) produce a localized melt pool that evolves into a global magma ocean; equilibration of the giant impactor core occurs at the melt-pool base ($P_{3}$), whereas small impacts (planetesimals) accreted during the magma-ocean lifetime equilibrate within the molten mantle at pressure $P_{4}$. }\label{Fig3}
\end{figure*}

\subsubsection{Equilibration pressure and extent of metal equilibration}
\paragraph{A) Early phase impacts.}
The early accretion phase is defined as $t {\leq} 5 \ \rm Myr$, coincident with the lifetime of the protoplanetary disk. During this period, protoplanets are subjected to intense heating by short-lived radionuclides \citep{Grimm1993,Tang_2012}, such as $^{26} \rm Al$ and $^{60} \rm Fe$ (see Fig. 4 of \citealt{McDonough_2020}). This heating is sufficient to sustain extensive or even global silicate melting, giving rise to magma oceans \citep{kleine_2009, rubie_2011}. In this regime, we assume the mantle to be fully molten, enabling efficient chemical homogenization \citep{Elkins_2008}. We also assume that after each collision, the bulk material of the impactor and target is fully mixed by the vigorous convection in this phase.\footnote{ Impactors and targets are referred to as small and large bodies during collisions hereafter.}
Global melting ensures that each impact effectively resets the chemical state, erasing the detailed record of previous equilibration events. Consequently, our model approximates early-stage differentiation by evaluating a single metal-silicate equilibration at the end of this molten stage, based on the planet's bulk composition at that time.

Metal-silicate equilibration during this stage is assumed to occur near the core-mantle boundary (CMB). Following previous studies, the equilibration pressure is parameterized as half the CMB pressure ($P_{\rm CMB}$), consistent with constraints inferred for Mars and other small terrestrial bodies \citep{dauphas_2011,Righter_2011,Rai_2013,brennan_2020}. This yields $P_1 {=} 112.06 \times M_{\mathrm{p}} + 0.37$, 
where $P_1$ is in GPa and $M_{\mathrm{p}}$ is the mass of the protoplanet at $t{=}5$ Myr (in Earth masses). These assumptions provide the pressure-temperature conditions used for metal-silicate equilibration during the early impact stage, as illustrated in Fig.~\ref{Fig3}A.

\paragraph{ B) Late phase, small impacts.}

After the dissipation of the gas disk ($t{>}5\ \rm Myr$), short-lived radiogenic heat sources are largely exhausted and protoplanets are assumed to have cooled sufficiently that long-lived global magma oceans are no longer sustained. The thermal state of the protoplanetary mantle is therefore assumed to be predominantly solid, except for localized melting induced by large impacts. 

At this stage, protoplanets that later evolve into Earth and Mars analogs typically reach masses exceeding $10^{-2}\,\rm M_{\oplus}$. We classify impacts into two types based on the impactor-to-total mass ratio, $\gamma = M_{\rm imp}/(M_{\rm imp}+M_{\rm tar})$. Small impacts are defined by $\gamma < 0.1$, whereas large impacts correspond to $\gamma \ge 0.1$.

For small impacts, the impact energy is assumed to be insufficient to generate large-scale surface melting or a global magma ocean on the target. Consequently, impact-induced melting remains localized and does not trigger mantle-wide mantle convection (see Fig.~\ref{Fig3}B). Under these conditions, we assume that $70\%$ of the impactor's metallic core, together with its entire silicate mantle, undergoes metal-silicate equilibration within the localized melt region generated in the target mantle. In this treatment, only the impactor-derived material participates in chemical equilibration, while the preexisting target mantle acts solely as a physical medium for the process and does not re-equilibrate itself. The remaining $30\%$ of the core descends through the target's mantle and directly merges with the target's core. 
Following previous studies \citep{rubie_2015,gu_2023}, we adopt the equilibration pressure for small impacts as $P_{2} {=} 0.6 P_{\rm CMB}$, where $P_{\rm CMB}$ is given by an empirical formula for the core-mantle boundary pressure, with details provided in Appendix B.

\paragraph{C)  Late phase, large impacts.}

In contrast, large impacts deliver sufficient energy to generate extensive melting in the target mantle. Such impacts are assumed to first produce a localized melt pool in the target mantle (left of Fig.~\ref{Fig3}C). During this stage, the metallic core of the large impactor equilibrates with silicate melt within the impact-generated melt pool. Following the scaling laws of \citet{Nakajima_2021}, the equilibration pressure is parameterized as $P_{3}=0.7P_{\rm b}$, where $P_{\rm b}$ is the pressure at the base of the melt pool. In this regime, $30\%$ of the impactor's metallic core \citep{gu_2023}, the entire impactor mantle, and a fraction $f_{\rm MP}$ of the target mantle are assumed to participate in metal-silicate equilibration. The value of $f_{\rm MP}$ is calculated self-consistently following the scaling laws of \citet{Nakajima_2021}.

The impact-generated melt pool is subsequently assumed to evolve into a global magma ocean (right of Fig.~\ref{Fig3}C), which progressively cools and crystallizes over a timescale of approximately 5 Myr \citep{Elkins_2008,dale_2025}. Smaller bodies accreted during this magma ocean stage descend through the molten mantle and progressively equilibrate with the silicate melt near the base of the magma ocean. The pressure at the base of the magma ocean is denoted by $P_{\rm MO}$ and is computed from the impact parameters recorded in the N-body simulation using the scaling laws of \citet{Nakajima_2021}. For simplicity, metal-silicate equilibration associated with these post impact accretion events is parameterized as $P_{4}=0.5P_{\rm MO}$ \citep{dale_2025}. In this stage, $70\%$ of impactor's metallic core, the entire impactor mantle, and a fraction $f_{\rm MO}=0.05$ of the target mantle are assumed to participate in metal-silicate equilibration \citep{dale_2025}. The equilibrated fractions adopted for each impact regime are summarized in Table~\ref{tab:equil_fraction}.

These equilibration fractions are motivated by impact simulations and scaling-law studies, which reflect incomplete metal emulsification and spatially limited mantle melting during impacts. Consistent with previous studies \citep{rubie_2015,fischer_2017,gu_2023}, we find that reasonable variations in these parameters can primarily affect absolute element abundances in planetary bodies. Nevertheless, the relative Earth-Mars compositional trends identified here remain robust.

\begin{table*}
\centering
\caption{Fractions of the impactor and target participating in metal-silicate equilibration for different impact regimes. }
\label{tab:equil_fraction}
\resizebox{0.95\textwidth}{!}{
\begin{tabular}{lcccc}
\hline
 Impact regime 
& Impactor's core 
& Impactor's mantle 
& Target's core 
& Target's mantle \\
\hline
A. early phase impact 
& 1.0 & 1.0 & 1.0 & 1.0 \\
B. late phase, small impact 
& 0.7 & 1.0 & 0.0 & 0.0 \\
C1. late phase, large impact 
& 0.3 & 1.0 & 0.0 & $f_{\mathrm{MP}}$ \\
C2. late phase, planetesimal accretion in post-large impact  
& 0.7 & 1.0 & 0.0 & $f_{\mathrm{MO}}$ \\
\hline
\end{tabular}
}
\tablefoot{For large impacts,  $f_{\mathrm{MP}}$ denotes the fraction of the target's  mantle involved in equilibration within the impact-generated melt pool, while  $f_{\mathrm{MO}}$ 
is the fraction of the basal magma-ocean mantle that equilibrates during subsequent planetesimal accretion.}
\end{table*}

\subsubsection{Equilibration temperature}

Following \cite{wade_2005} and \cite{rubie_2015}, the equilibration temperature $T_{\rm eq}$ is determined from its corresponding pressure $P_{\rm eq}$, assuming a value between the peridotite solidus and liquidus. Such temperatures are parameterized using the experimentally constrained peridotite melting relations of \cite{rubie_2015}:
\begin{equation}
\begin{cases}
\text{if } P_{\rm eq} < 24\ \text{GPa}, \\
T_{\rm eq} = 1874 + 55.43P_{\rm eq} - 1.74P_{\rm eq}^2 + 0.0193P_{\rm eq}^3, \\ 
\text{if } P_{\rm eq} \ge 24\ \text{GPa}, \\
T_{\rm eq} = 1249 + 58.28P_{\rm eq} - 0.395P_{\rm eq}^2 + 0.0011P_{\rm eq}^3, 

\end{cases}
\end{equation}
with $T_{\rm eq}$ in Kelvin and $P_{\rm eq}$ in GPa.

\subsection{Integrated workflow: AccreDiff}
To unify the framework described above, we developed AccreDiff (accretion plus differentiation), a Python-based, modular and extensible workflow designed to couple dynamical accretion histories with impact-driven core-mantle differentiation calculations. In this study, accretion dynamics (including collision sequences, mass growth, and source-region mixing) are taken as inputs from N-body simulations performed with the GPU-accelerated code GENGA \citep{grimm_2014a,grimm_2022}. AccreDiff then computes the corresponding chemical evolution through multistage metal-silicate equilibration and element partitioning. The AccreDiff framework will be made publicly available in a forthcoming publication (Kong et al., in prep.).

\section{Results}
In this section, we present the results of our simulations. Owing to the stochastic nature of N-body simulations \citep{wisdom_1987, laskar_1997a}, an exact Earth-Mars analog system is difficult to reproduce in every realization. Out of a total of 28 $N$-body simulations, seven runs produced systems containing both an Earth analog and a Mars analog, defined as inner planet with mass $M_{\rm p}{=}0.7{-}1.0 \rm \ M_{\oplus}$ and semimajor axis $a_{\rm p}{=}0.7{-}1.3$ AU, together with an outer planet with $M_{\rm p}{=}0.05{-}0.15 \rm \ M_{\oplus}$ and $a_{\rm p}{=}1.3{-}2.3$ AU. Among these seven systems, one reproduces an Earth-Mars pair with no additional intermediate-mass bodies between the two planets, while the remaining six contain low-mass bodies situated between the Earth and Mars analogs. All seven systems were retained for the geochemical analysis below, as no systematic difference exists between these two groups. The orbital architectures of the above seven systems are summarized in Appendix C. The present work provides a large-scale numerical investigation of the coupled dynamical and geochemical evolution of terrestrial planets within the narrow-ring accretion scenario; each simulation requires approximately 25 days of dedicated runtime on a modern NVIDIA A800 GPU.

In Sect. 3.1 we describe the accretion history and source materials of the Earth and Mars analogs. In Sect. 3.2 we discuss the partitioning of elements between the cores and mantles of these planets. In Sect. 3.3 we summarize the bulk properties of the resulting planets, and we explain why Earth and Mars have different geochemical signatures in Sect. 3.4.

\subsection{Accretion history and sources} 

\begin{figure*}
\sidecaption
    \includegraphics[width=12cm]{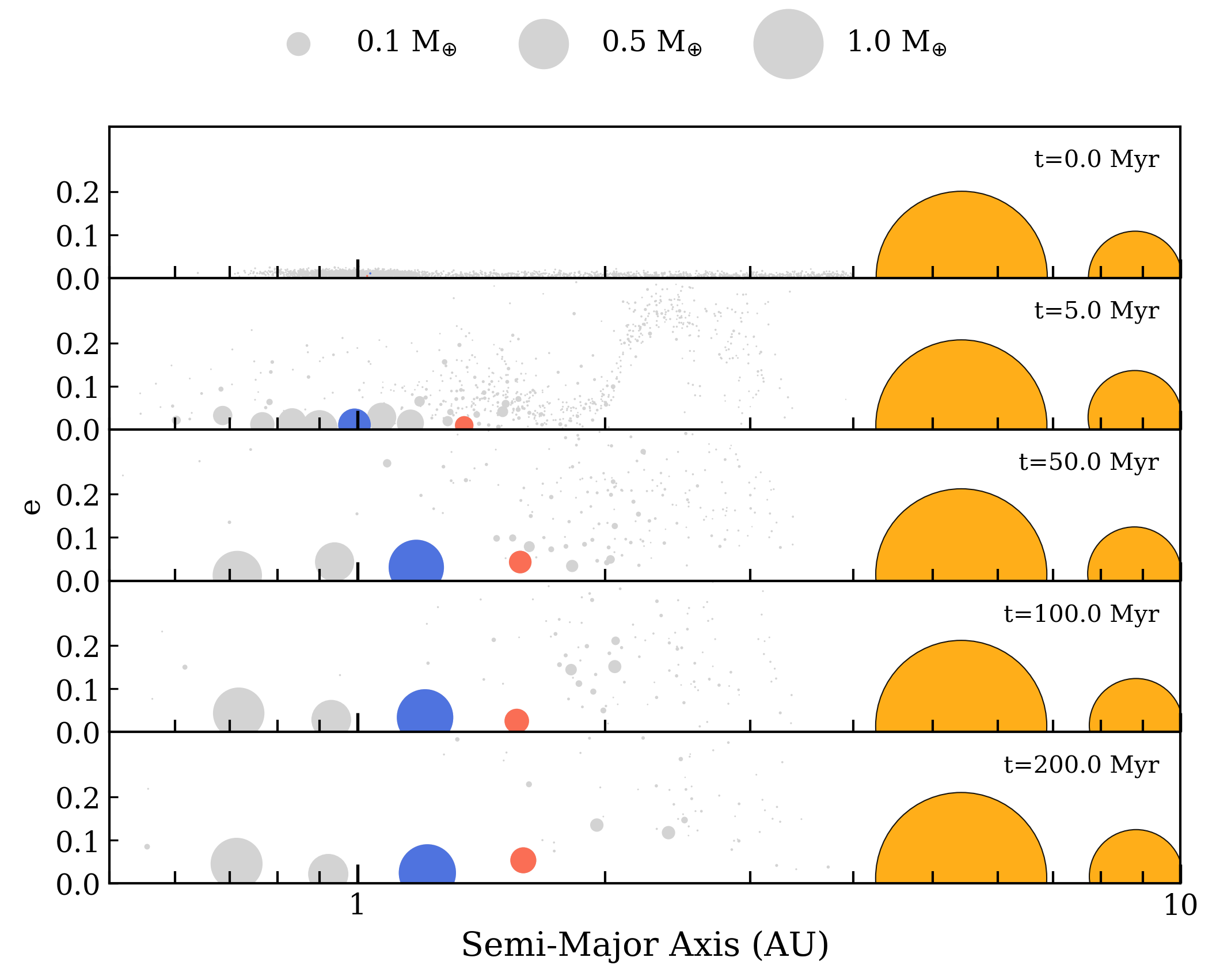}
    \caption{Snapshots of the dynamical evolution of the system shown in the semi-major axis-eccentricity plane from 0 to 200 Myr. Light-gray points indicate planetesimals; orange points mark Jupiter and Saturn; blue and red points show the Earth and Mars analogs, with sizes scaled to mass. Earth analogs, which achieve a final mass of approximately 0.71 $\rm M_{\oplus}$, experience rapid growth within the dense annulus. In contrast, the Mars analog embryo, which reaches a final mass of about 0.14 $\rm M_{\oplus}$, is scattered outward at an early stage and retains a low mass within a depleted area.}
    \label{Fig4}
\end{figure*}

Fig.~\ref{Fig4} shows the mass and orbital evolution of a Earth-Mars pair analog system from our simulations. Light-gray points represent the planetesimal disk, while blue, red, and orange circles denote the analogs of Earth, Mars and two gas giants (Jupiter and Saturn), respectively, with symbol sizes scaled by mass. Planetesimals in the ring region experience the highest solid surface density, undergo frequent collisions, and rapidly grow into larger bodies. The proto-Earth remains near the center of the ring. In contrast, the proto-Mars is scattered outward relatively early ($t {<} 5\ \mathrm{Myr}$). Once displaced beyond the dense ring, the proto-Mars experiences a much lower planetesimal surface density, and its accretion rate slows substantially, leaving it as a stranded low-mass body. This outcome is consistent with previous dynamical studies \citep{hansen_2009a,izidoro_2022a,nesvorny_2021b}.

\begin{figure}
\centering
\sidecaption
\includegraphics[width=\linewidth]{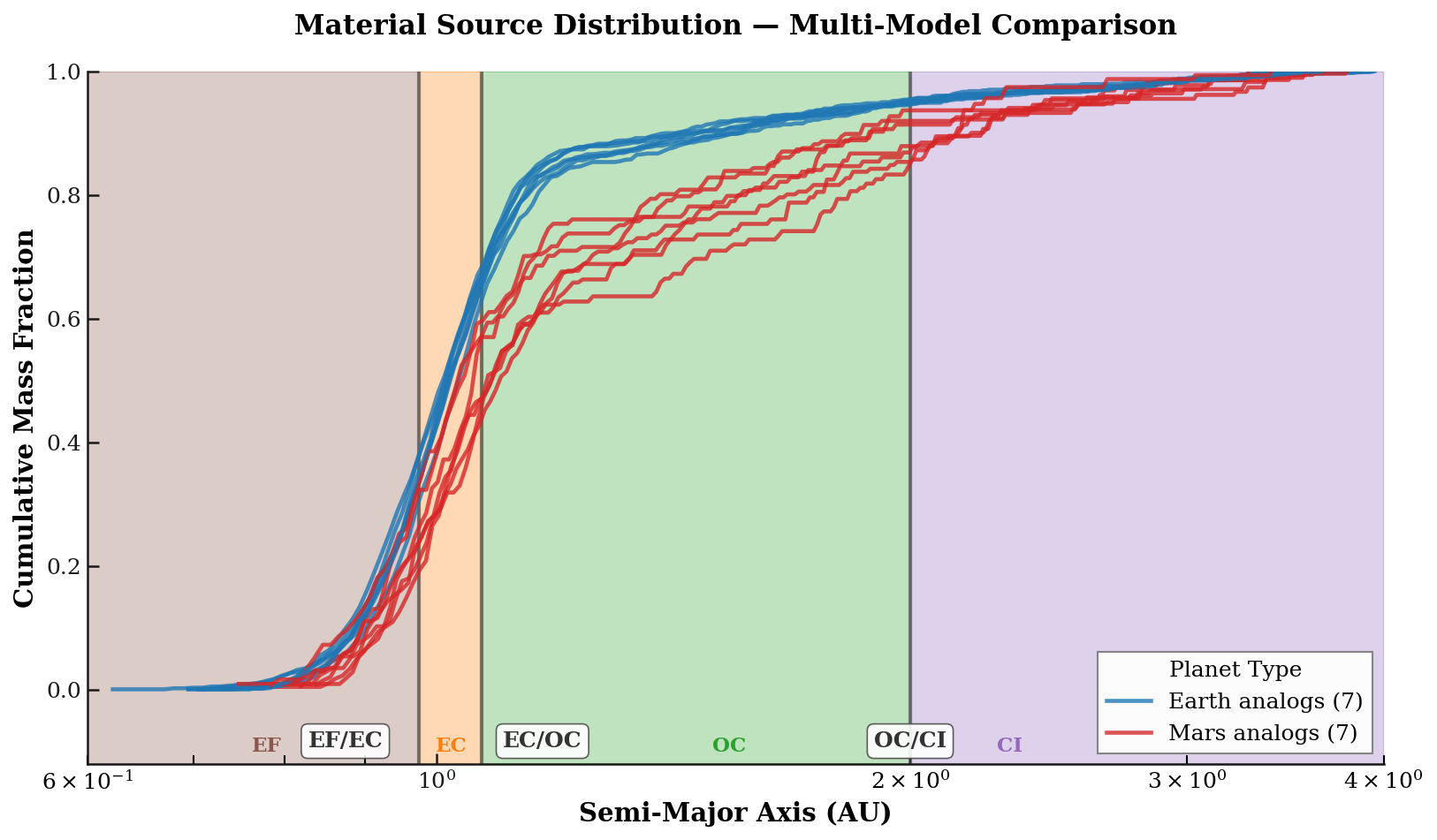}
\caption{Cumulative mass-fraction distributions of accreted material as a function of initial semi-major axis for Earth analogs (blue) and Mars analogs (red). Shaded regions indicate the adopted compositional reservoirs: EF ($a < 0.9747$ AU), EC ($0.9747\leq a < 1.0674$ AU), OC ($1.0674\leq a < 2.0$ AU), and CI ($a\geq 2.0$ AU). Vertical black lines mark the boundaries between adjacent reservoirs. Earth analogs preferentially accrete inner-disk material (EF-EC), whereas Mars analogs show enhanced contributions from outer-disk material (OC-CI).}\label{Fig5}
\end{figure}

Fig.~\ref{Fig5} depicts the cumulative mass distributions of accreted material as a function of the initial semi-major axis for Earth and Mars analogs. The shaded regions denote the adopted EF, EC, OC and CI compositional reservoirs. Earth analogs exhibit steep cumulative increases within the EF-EC regions, indicating that most of their accreted material originates from reduced inner-disk reservoirs. In contrast, Mars analogs display broader cumulative distributions extending into the OC-CI regions, reflecting enhanced contributions from more oxidized outer-disk material. The systematic difference in radial material provenance naturally produces relatively reduced bulk compositions for Earth analogs and more oxidized bulk compositions for Mars analogs.

\begin{figure}
\centering
\includegraphics[width=0.95\linewidth]{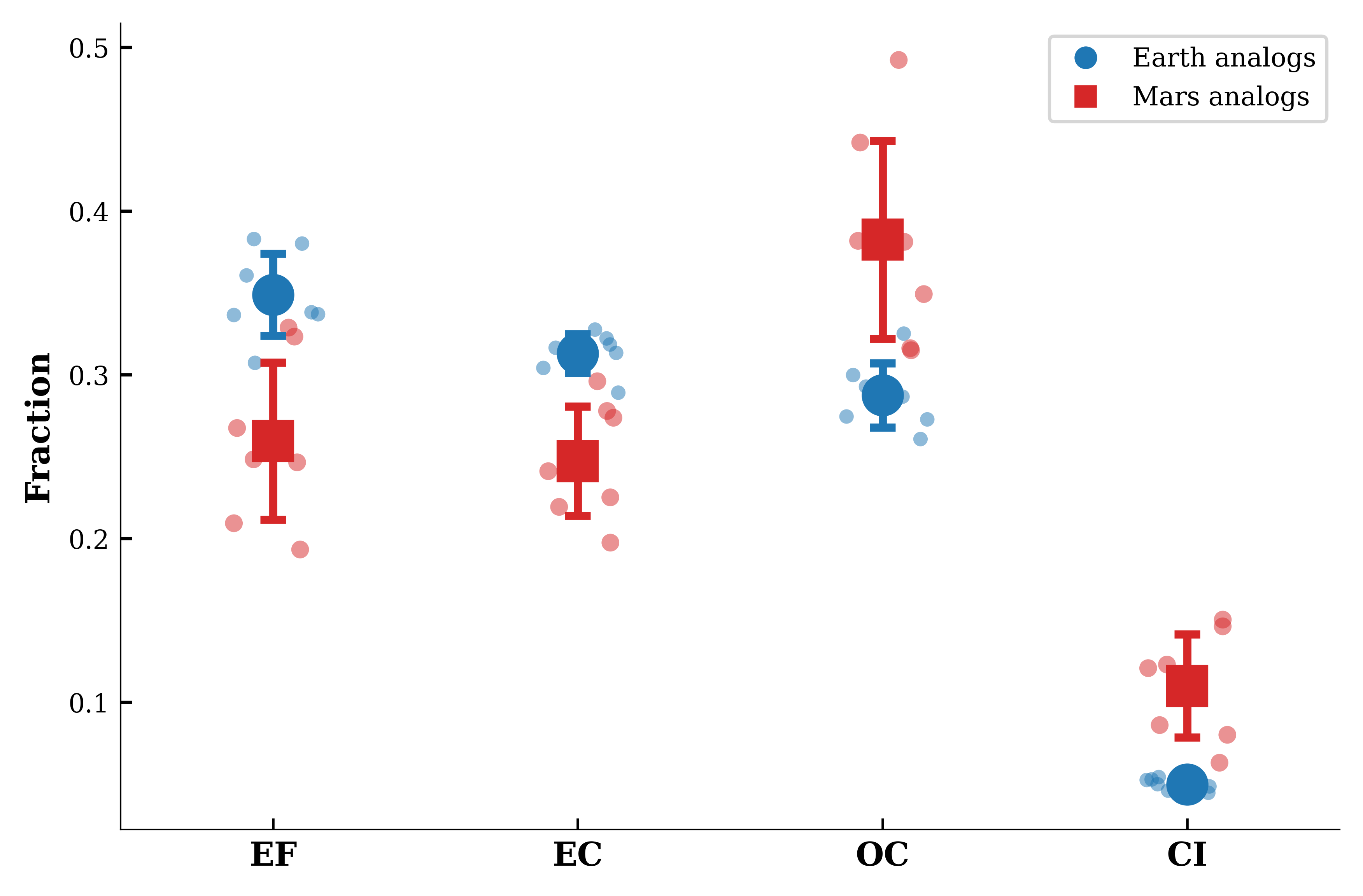}
\caption{Ensemble-averaged accretion source fractions of EF, EC, OC, and CI material for Earth analogs (blue) and Mars analogs (red) across all qualifying systems. Dots show individual simulations, and error bars indicate the mean $\pm 1\sigma$. Earth analogs preferentially accrete reduced inner-disk components (EF-EC), whereas Mars analogs exhibit enhanced contributions from OC material; CI fractions remain minor in both cases.}\label{Fig6}
\end{figure}

To assess the statistical robustness of the radial provenance trend shown in Fig.~\ref{Fig5}, we analyze the full set of seven successful simulations. As shown in Fig.~\ref{Fig6}, the ensemble results closely reproduce the trends seen in the representative system: Earth analogs consistently accrete higher fractions of reduced EF and EC material, whereas Mars analogs exhibit elevated OC contributions; CI remains a minor component for both planets. The detailed accretion fractions of each end‑member for these systems are listed in Appendix C.

\subsection{Core-mantle chemical differentiation}
\begin{figure}
\centering
\sidecaption
\includegraphics[width=\linewidth]{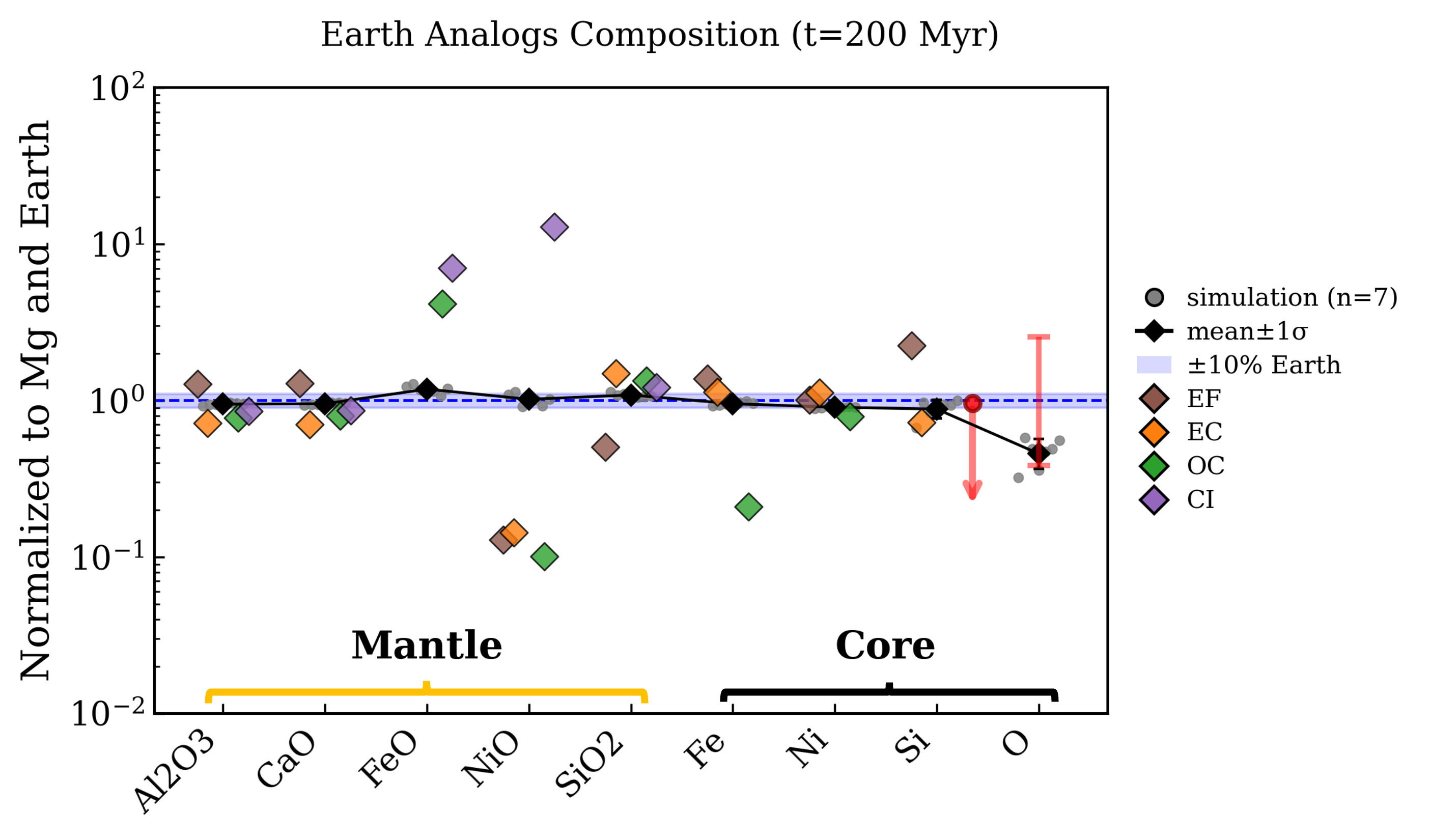}
\caption{Mantle (oxide) and core (elements) compositions of Earth analogs at 200 Myr, normalized to Mg and further normalized to Earth reference atomic ratios. Gray circles show individual simulations (n=7). Black diamonds with error bars indicate the mean values with a $\pm 1\sigma$ spread. Colored diamonds denote the compositions of the meteoritic endmembers used in the accretion-composition tagging scheme (EF, EC, OC, and CI). The dashed blue lines and shaded band denote the Earth reference values ($\pm 10 \%$). Red markers indicate the estimated uncertainty ranges of Si and O abundances in the Earth's core from \citet{Hirose_2021}}\label{Fig7}
\end{figure}

\begin{figure}
\centering
\sidecaption
\includegraphics[width=\linewidth]{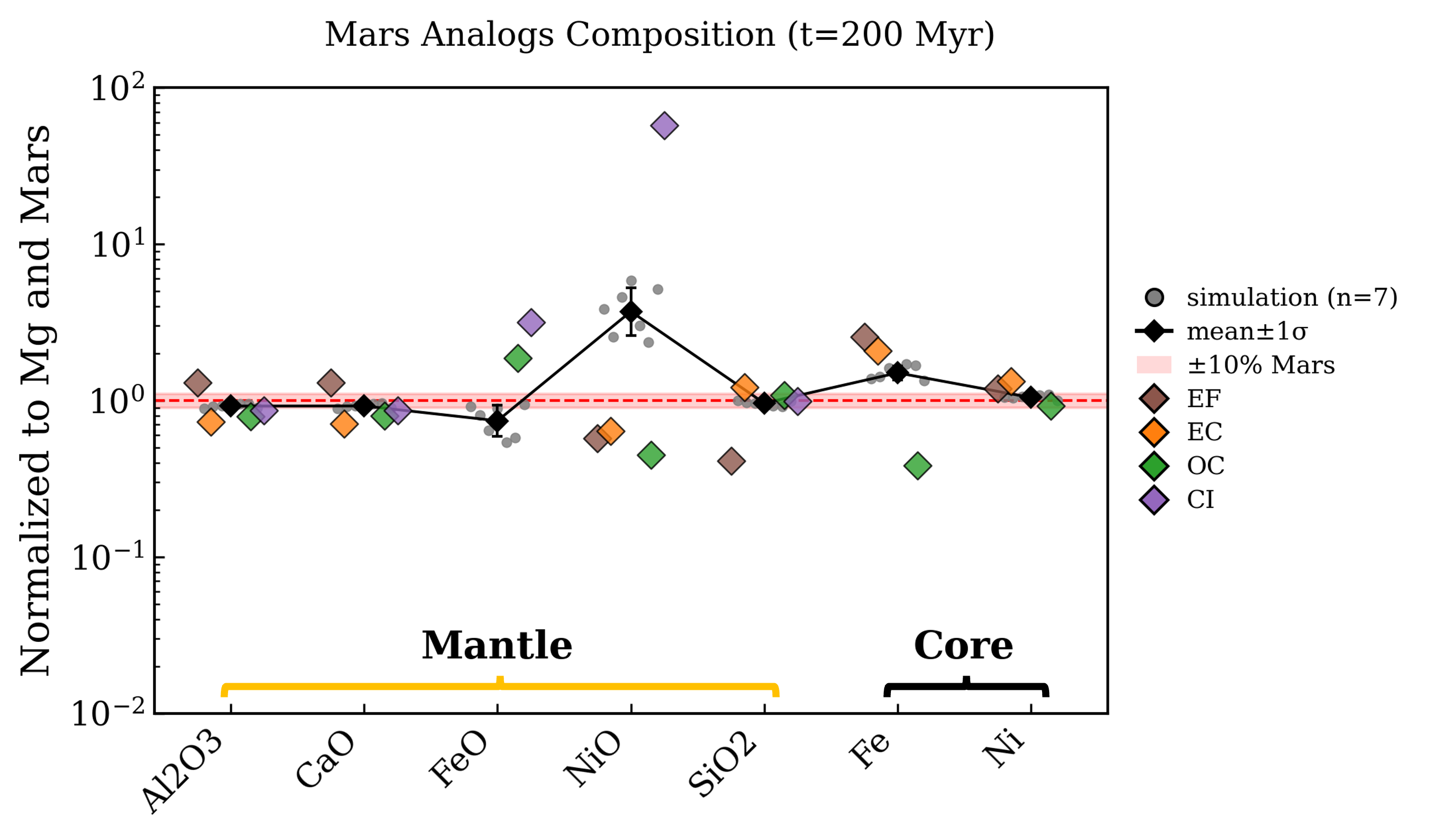}
\caption{Mantle (oxide) and core (elements) compositions of Mars analogs at 200 Myr, normalized to Mg and further normalized to Mars reference atomic ratios. Gray circles show individual simulations (n=7). Black diamonds with error bars indicate the mean values with a $\pm 1\sigma$ spread. Colored diamonds denote the compositions of the meteoritic endmembers used in the accretion-composition tagging scheme (EF, EC, OC, and CI). The dashed red lines and shaded band denote the Mars reference values ($\pm 10 \%$).}\label{Fig8}
\end{figure}

Figs.~\ref{Fig7} and \ref{Fig8} present the modeled mantle and core compositions of Earth and Mars analogs, expressed in atomic proportions (atomic \%) , normalized to Mg and to their respective planetary reference compositions. For reference, the same compositions expressed in weight percent, as commonly reported in the geochemical literature, are provided in Appendix D. 

For Earth analogs shown in Fig.~\ref{Fig7}, mantle lithophile oxides (Al$_2$O$_3$, CaO) and SiO$_2$, NiO, broadly reproduce bulk silicate Earth (BSE) values, with small inter-system variability (Table~\ref{tab:comp_vertical}).  
By contrast, mantle FeO is systematically slightly higher than the BSE value. The origin of these offsets is discussed in Section~4.3. Core Fe, Ni and Si abundances remain broadly consistent with the reference values. In contrast, core O is systematically depleted in most Earth analogs (Fig.~\ref{Fig7}; Table~\ref{tab:comp_vertical}). These trends likely reflect the strong sensitivity of light-element partitioning to equilibration $P$--$T$ conditions and bulk redox state. We note that light-element abundances in planetary cores remain subject to substantial observational and modeling uncertainties. For example, the estimated O and Si contents of Earth's core span broad ranges of 0.8$\sim$5.4 wt\% and 0$\sim$4.0 wt\%, respectively \citep{Hirose_2021}, as indicate by the red markers in Fig.~\ref{Fig7}. Therefore, core light-element abundances should not be interpreted as tightly constrained benchmarks \citep{Hirose_2021}.

As shown in Fig.~\ref{Fig8}, Mars analogs broadly reproduce the reference values of Al$_2$O$_3$, CaO and SiO$_2$ within the $\pm 10\%$ range. Mantle FeO is slightly lower than the reference value, whereas NiO is consistently enriched and exhibits relatively large inter-system dispersion. On the core side, Mars analogs show systematically elevated Fe abundances compared to the reference value, while Ni remains broadly consistent with the martian reference composition.

\begin{table*}
\centering
\caption{Mean compositions of Earth analogs and Mars analogs at 200 Myr, normalized to Mg (by atomic).}
\label{tab:comp_vertical}
\setlength{\tabcolsep}{20pt}
\renewcommand{\arraystretch}{1.2}
\begin{tabular}{llll}
\hline
\multicolumn{4}{c}{\textbf{Mantle compositions (oxides)}} \\
\hline
Oxide & Earth analogs & Mars analogs & Reference \\
\hline
Al$_2$O$_3$
& 0.0440 $\pm$ 0.0006
& 0.0422 $\pm$ 0.0012
& Earth: 0.0464 / Mars: 0.0458 \\

CaO
& 0.0641 $\pm$ 0.0010
& 0.0616 $\pm$ 0.0016
& Earth: 0.0673 / Mars: 0.0668 \\

FeO
& 0.1417 $\pm$ 0.0082
& 0.2018 $\pm$ 0.0452
& Earth: 0.1196 / Mars: 0.2660 \\

NiO
& 0.0036 $\pm$ 0.0003
& 0.0031 $\pm$ 0.0011
& Earth: 0.0036 / Mars: 0.0008 \\

SiO$_2$
& 0.8640 $\pm$ 0.0227
& 0.9467 $\pm$ 0.0300
& Earth: 0.7968 / Mars: 0.9846 \\

\hline
\multicolumn{4}{c}{\textbf{Core compositions (elements)}} \\
\hline
Element & Earth analogs & Mars analogs & Reference \\
\hline
Fe
& 0.7562 $\pm$ 0.0174
& 0.6513 $\pm$ 0.0674
& Earth: 0.7931 / Mars: 0.4319 \\

Ni
& 0.0408 $\pm$ 0.0004
& 0.0400 $\pm$ 0.0013
& Earth: 0.0450 / Mars: 0.0383 \\

Si
& 0.0654 $\pm$ 0.0083
& 0.0133 $\pm$ 0.0063
& Earth: 0.0738 / Mars: -- \\

O
& 0.0302 $\pm$ 0.0062
& 0.0053 $\pm$ 0.0013
& Earth: 0.0648 / Mars: 0.0986 \\
\hline
\end{tabular}
\tablefoot{Mantle compositions are reported as oxides, whereas core compositions are reported as elemental abundances.
Uncertainties denote $\pm 1\sigma$ across the selected systems ($n=7$).}
\end{table*}

\subsection{Bulk planetary properties}
\begin{figure}
\centering
\includegraphics[width=0.95\linewidth]{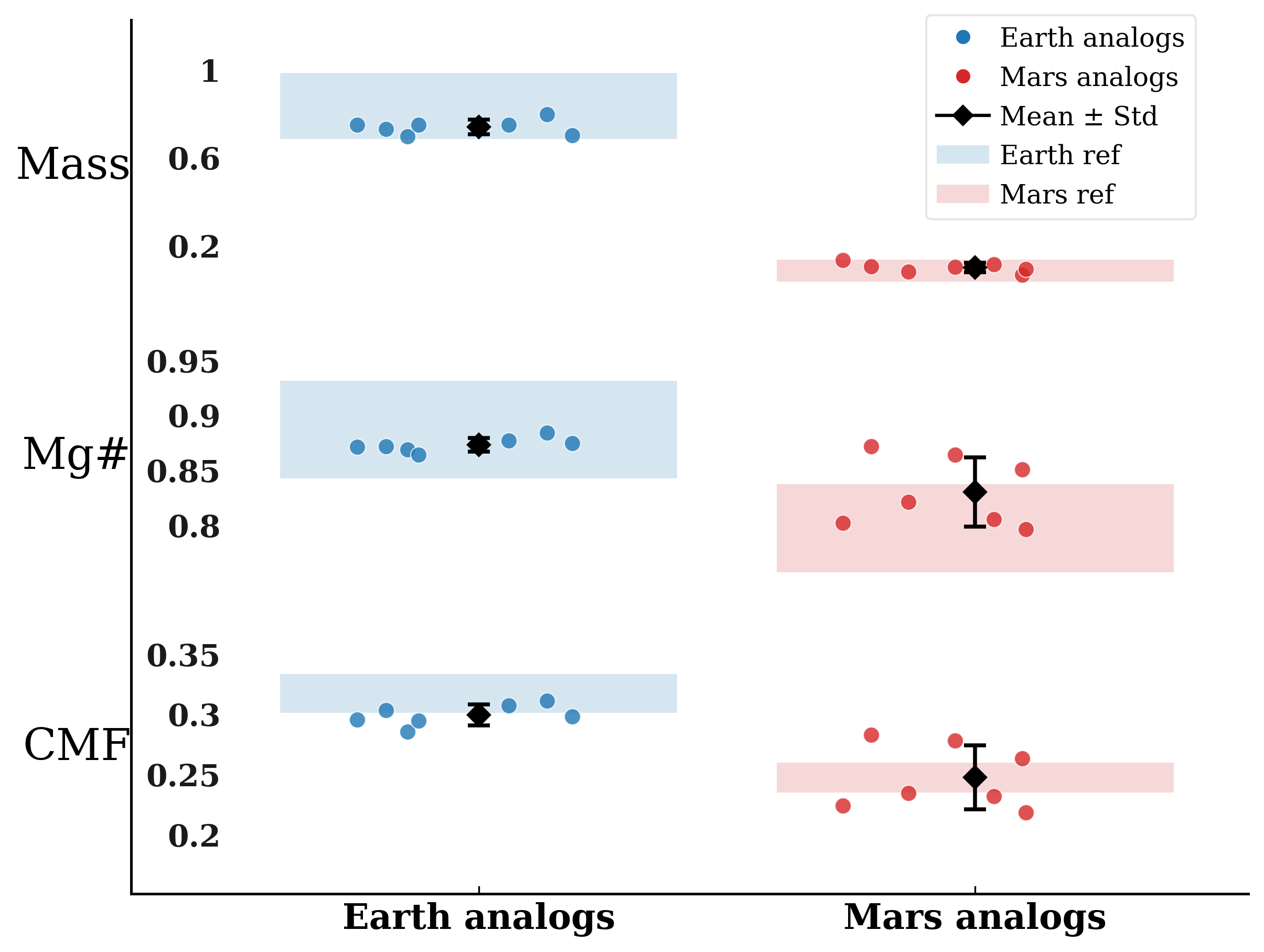}
\caption{ Comparison of bulk properties between Earth analogs and Mars analogs at 200 Myr, including mass in units of Earth mass, mantle Mg\#, and CMF. Individual simulations are shown together with mean $\pm 1\sigma$. Shaded regions denote the adopted reference intervals: For Earth, Mass=$0.7{-}1.0 M_{\oplus}$, Mg\#=$0.85{-}0.93$, and CMF=$0.30{-}0.34$; for Mars, Mass=$0.05{-}0.15 M_{\oplus}$, Mg\# =$0.76{-}0.84$, and CMF=$0.24{-}0.26$.}\label{Fig9}
\end{figure}

Fig.~\ref{Fig9} illustrates key bulk planetary properties for Earth and Mars analogs, including final mass, mantle Mg\# (the molar Mg/(Mg+Fe) ratio in the mantle), and core mass fraction (CMF; the ratio of core mass to total planetary mass). Together, these metrics provide combined constraints on planetary accretion and core-mantle differentiation, and serve as key benchmarks for model evaluation.

Despite stochastic variations in individual accretion histories, the simulated planets exhibit a clear mass dichotomy between the two planet classes. Earth analogs systematically reach larger final masses, with a mean value of $\sim 0.75\ M_{\oplus}$, whereas Mars analogs remain substantially smaller, with a mean mass of $\sim 0.11\ M_{\oplus}$. Both populations fall within their adopted reference intervals and display relatively small inter-system dispersion.

Mantle Mg\# is a widely used indicator of planetary redox state, which is defined as 
\begin{equation}
{\rm Mg\#} = \frac{\rm Mg}{\rm Mg + Fe},
\end{equation}
where Mg and Fe denote the molar abundances of these two elements in the silicate mantle.  A lower Mg\# corresponds to a more oxidized mantle with a higher proportion of Fe retained as FeO, whereas a higher Mg\# reflects a more reduced mantle. In our simulations, Earth analogs remain broadly consistent with the terrestrial reference range and generally exhibit higher Mg\# values than Mars analogs. In contrast, Mars analogs show lower Mg\#, indicating relatively more oxidized mantle compositions. Although moderate overlap exists between the two populations, the overall offset in Mg\# reflects systematic differences in bulk redox state established by their distinct radial accretion pathways.

The CMF further reinforces this dichotomy and is defined as
\begin{equation}
{\rm CMF} = \frac{M_{\rm core}}{M_{\rm tot}},
\end{equation}
where $M_{\rm core}$ and $M_{\rm tot}$ are the core and total planet masses. Earth analogs exhibit CMF values $\sim$ 0.30 close to the reference of 0.32, indicating efficient segregation of metallic iron into the core. 

Mars analogs have tend to have smaller cores (${\sim} 0.25$). This trend is consistent with their more oxidized accretion histories, which promote Fe retention as FeO in the mantle and reduce the amount of metallic Fe available for core formation. 
The slightly elevated CMF of the Mars analogs should be interpreted with caution. Recent revisions of the martian mantle FeO content, from $18 \%$ to $14 \%$, imply a lower mantle iron fraction. Under a pure-iron-core model, the martian CMF is approximately 0.25 \citep{Khan_2022,Guimond_2023}. When accounting for sulfur, a key light element expected to be present in the martian core as Fe-S alloys, the martian CMF can be as low as 18\%. Since sulfur is not included in the present model, our CMF estimates represent sulfur-free end-member outcomes. The small offset from the martian reference likely reflects both this compositional simplification and the inherent uncertainty in the martian CMF. Taken together, Earth and Mars analogs occupy distinct regions in the parameter space (mass, Mg\#, CMF), providing joint constraints on the coupled processes of accretion and core-mantle differentiation.

\subsection{Compositional difference of Earth and Mars }
The compositional differences between Earth and Mars analogs are governed by the coupled effects of accretion dynamics, the radial redox structure of solid reservoirs in the protoplanetary disk, and impact-driven core-mantle differentiation. Although both analogs originate near 1 AU, their distinct dynamical histories lead them to sample systematically different chemical reservoirs. Earth analogs preferentially accrete reduced inner-disk material (EF and EC), whereas Mars analogs acquire a larger fraction of more oxidized material (OC and CI) from farther out in the disk \citep{righter_2020,grewal_2021,mcdonough_2021a,stagno_2021}. These contrasting accretionary sources establish distinct bulk redox states of the planets.

We note that while accretion dynamics and the disk's radial redox structure determine what is delivered to the forming planets, impact-driven equilibration governs how the elements are distributed between the metallic core and the silicate mantle. Importantly, although Mg\# and core mass fraction (CMF) are often discussed together, they are not equivalent: CMF is a structural metric, whereas Mg\# is a compositional metric constrained independently. However, they remain physically coupled through Fe redistribution during differentiation, such that more efficient Fe segregation into the core increases CMF while lowering mantle FeO and raising Mg\#. Accordingly, Earth analogs accrete a larger fraction of reduced inner-disk material, enabling a greater proportion of their bulk Fe to be incorporated into the core, resulting in higher Mg\# and larger CMFs. By contrast, Mars analogs accrete relatively more oxidized material, which thermodynamically favors the conversion of Fe into FeO retained in the mantle and thus limits core growth, producing lower Mg\# and smaller CMFs \citep{righter_1996,malavergne_2007,boujibar_2014}.

\section{Discussion}
\subsection{Comparison with previous models}
Several studies have demonstrated that coupling accretion dynamics with metal-silicate differentiation can successfully reproduce the composition of the bulk Earth (e.g., \citealt{rubie_2015,fischer_2017,gu_2023,dale_2025}). In particular, \cite{dale_2025} combined N-body simulations with differentiation calculations using multi-compositional reservoirs (EC, OC, CI, and a refractory-rich inner-disk component) and showed that key characteristics of Earth can be reproduced.

Our model shares several similarities with \cite{dale_2025}, including the narrow-ring accretion scenario. This configuration is motivated by dynamical studies showing that narrow-ring models can simultaneously reproduce the orbital architecture and growth timescales of the terrestrial planets when gas disk evolution is considered (e.g., \citealt{nesvorny_2021b,Woo_2023,woo_2024}). In addition, we also adopt the above four-reservoir compositional structure. This configuration allows lithophile element ratios such as Al/Mg and Ca/Mg to vary among forming planets through differences in accretional provenance. By contrast, models that prescribe identical initial silicate compositions for all embryos (e.g., CI-based compositions with varying oxygen fugacity \citealt{rubie_2011,rubie_2015}) would produce nearly identical lithophile element ratios for Earth and Mars analogs, inconsistent with the observed differences between their mantles (see Fig.~\ref{Fig2}).

Despite these similarities in the initial setup, the goal of the present work differs from previous studies. Rather than tuning model parameters to reproduce the bulk Earth composition, the focus of our investigation is whether the Earth-Mars compositional difference can arise within a single dynamical framework. In addition, unlike previous studies such as \citet{dale_2025}, which mainly focused on the post-gas-disk giant-impact stage, the present work explicitly includes the gas-disk phase (0{-}5 Myr), which is likely critical for the early growth and evolution of Mars-sized bodies. In this sense, the present study can serve as an independent exploration that extends the dynamical-geochemical coupling framework to a broader context. Furthermore, \citet{Nathan_2023} noted that, whitin the Grand Tack framework, the rapid and stochastic growth of Mar analogs make them difficult to reproduce an oxidation gradient compatible with both Earth and Mars. By contrast, the higher resolution of our narrow ring simulations produces smoother accretion histories for Mars analogs (see Fig.~\ref{Fig5}), allowing the divergent evolution of Earth and Mars analogs to emerge within the same framework.

Although the adopted initial radial compositional gradient is prescribed, recent dynamical studies indicate that such structures can emerge during the early evolution of the protoplanetary disk. In particular, \citet{Goldberg_2026} proposed that early dynamical evolution may preferentially redistribute planetesimals originating at different heliocentric distances, thereby enhancing the delivery of outer-disk material to the Mars-forming region. In addition, recent high-resolution simulations by \citet{Shuai2025} showed that sweeping secular resonances associated with gas-disk dissipation can drive substantial inward transport and dynamical mixing of planetesimals, particularly for bodies initially located beyond $\sim$ 2 AU. These works provide a physically plausible basis for the radial oxidation gradient adopted in the present work and support the hypothesis that Earth and Mars may sample compositionally distinct reservoirs even within a narrow ring accretion model. Finally, while our work focuses on the chemical compositions of major elements, future studies should also integrate isotopic constraints into a consistent framework (e.g., \citealt{Dauphas_2017,Shuai2023}).

\subsection{Motivation for the hypothetical EF component}
Fig.~\ref{Fig2} shows that both Earth and Mars are systematically enriched in Al and Ca relative to EC, OC and CI chondrites. Because Al and Ca are refractory lithophile elements concentrated in the mantle, these elevated abundances cannot be reproduced by mixing known meteoritic reservoirs alone. If only EC-OC-CI mixtures are considered, the resulting bulk compositions remain systematically depleted in Al and Ca. Even in this case, the fitting solution is naturally driven toward CI-rich mixtures, because CI chondrites are the most Al- and Ca-rich end-members EC, OC and CI. However, increasing the CI contribution to match Al and Ca would simultaneously introduce excessive volatile-rich and oxidized material, leading to mantle compositions that are difficult to reconcile with independent geochemical constraints. This discrepancy implies an additional inner-disk component that is absent from the currently sampled meteoritic groups. We therefore introduce a hypothetical EF component enriched in refractory elements, particular Al and Ca. Similar refractory-rich inner Solar System reservoirs have been invoked to reconcile terrestrial-planet compositions with cosmochemical constraints, for example CAI-like material or other high-temperature condensates preferentially retained at small heliocentric distances \citep{moynier_2015,mezger_2020}.

In our model, EF material is confined to the disk region interior to 1 AU, consistent with the location of the silicate sublimation line (e.g., forsterite) in the early protoplanetary disk \citep{Larimer1970, Morbidelli_2022}. Interior to this line, high disk temperature favors the formation of Ca-, Mg- and  Al-rich, Si-depleted condensates \citep{Morrison_2020,Ebel_2021,guimond_2024}. Although EF does not belong to any known meteorite group, it serves as a physically motivated inner-disk end‑member that helps to reproduce the refractory enrichment of Earth and Mars while remaining broadly consistent with other cosmochemical constraints \citep{Dauphas2015}.

\subsection{Impact-controlled equilibration and light-element in cores}
Light-element incorporation into planetary cores provides a key link between impact conditions, redox evolution, and the resulting core-mantle compositions. Seismological constraints indicate that Earth's core is less dense than a pure Fe-Ni alloy, implying the incorporation of a $\sim$5{-}10\% mass fraction of light elements into the core \citep{Badro_2014,Hirose_2021}. Among the candidate light elements, Si (and to a lesser extent O) is expected to become increasingly siderophile at high pressures and temperatures, as demonstrated by metal-silicate partitioning experiments and thermodynamic models \citep{wade_2005,wood_2008}.

In our model, these two process exert opposite effects on mantle FeO abundance. The process of Si incorporation into the core occurs through reactions such as $\mathrm{SiO_{2}(mantle)}+2\mathrm{Fe(core)}\rightarrow \mathrm{Si(core)}+2\mathrm{FeO(mantle)}$(Reaction A), which increase mantle FeO, whereas O partitioning into the core effectively consumes FeO through reactions of the form $\mathrm{FeO(mantle)\rightarrow \mathrm{Fe(core)}+\mathrm{O(core)}}$(Reaction B), thereby reducing the FeO content of the silicate mantle \citep{rubie_2011,rubie_2015}. This competition is reflected in our Earth analogs Fig~\ref{Fig7}. The modeled mantles exhibit FeO contents slightly higher than the bulk silicate Earth value, suggesting that Reaction A is slightly stronger than Reaction B. At the same time, the modeled core Si and O abundances remain within the broad uncertainty ranges inferred for the Earth's core. For example, the estimated O and Si contents of the Earth's core span broad ranges of 0.8$\sim$ 5.4 wt\% and 0$\sim$ 4 wt\%, respectively \citep{Hirose_2021}, as indicated by the red markers in Fig.~\ref{Fig7}. Therefore, the present results do not require additional parameter tuning to exactly reproduce the nominal Si and O abundances of the Earth's core.

Although the precise core light-element inventory remains uncertain, these differences mainly affect the absolute abundances of Si and O, rather than the first-order redox evolution of the planets. The simulations still successfully reproduce the systematic Earth-Mars contrast in mantle Mg\# and CMF within a self-consistent N-body accretion framework, indicating that radial provenance and bulk accretional redox state exert the primary control on planetary differentiation.

\subsection{Elevated mantle NiO in Mars analogs}
Although the mantle NiO abundances of Earth analogs are broadly consistent with the reference values, Mars analogs systematically exhibit elevated mantle NiO relative to the martian composition (Fig.~\ref{Fig8}). A likely explanation for the elevated martian NiO is the absence of sulfur in the present model. Experimental and thermodynamic studies indicate that sulfur enhances the siderophile behavior of Ni, driving its partitioning into metallic liquids (e.g., \citealt{Gaetani_1997,righter_2003,boujibar_2014,mahan_2021,loroch_2024}). Because Mars is expected to possess a sulfur-rich core, neglecting sulfur likely causes excessive retention of Ni within the silicate mantle, leading to systematically elevated mantle NiO in the Mars analogs.

Importantly, this discrepancy does not affect the main conclusions of our study. The Earth-Mars compositional difference, as reflected in FeO, Mg\# and core mass fraction (CMF), is primarily governed by bulk redox state, accretional provenance, and equilibration $P$--$T$ conditions. NiO thus serves as a sensitive tracer of secondary processes not yet included in the model, such as sulfur chemistry and high-temperature volatility. The incorporation of these effects represents a promising direction for future model refinement.

\subsection{Model limitations}

Here we outline several simplifications made in our model. First, while the chosen dynamical model captures key aspects of terrestrial planet formation and reliably produces Earth-Mars analogs, it represents only one possible pathway. Alternative models such as Grand Tack, pebble accretion, early instability, sweeping secular resoannce \citep{Thommes2008, walsh_2011b,O_Brien_2014,Clement_2017,johansen_2021,Liu_2022} could sample different accretion histories, impactor frequencies, and chemical reservoirs, potentially altering core-mantle differentiation outcomes. A systematic comparison across formation models is an essential next step.

Second, core formation is treated as an S-free end-member. Sulfur, which is expected to influence Fe partitioning during the Martian core formation \citep{Mezger_2013,Rai_2013}, is not included in the present model. Therefore, the inferred core compositions of Mars analogs should be interpreted as first-order outcomes of metal-silicate equilibration in a sulfur-free system. Incorporating sulfur will be explored in future work and is expected to influence metal-silicate partitioning, late sulfide addition to the core \citep{Neill_1991,Zhang_2024}, the behavior of moderately siderophile elements, and the absolute value of CMF. This is especially true for Mars analogs, which likely have high S abundance \citep{Fei_1995,Fei_1997,Khan_2022} and would preserve the relative trends identified in this study \citep{righter_2003,righter_2020,lv_2022c}. Furthermore sulfur is significantly depleted in the bulk Earth, so Earth's mantle FeO is not expected to be strongly affected by this simplification.

Third, volatile loss and redistribution during high-temperature stages are not explicitly modeled. Such processes may modify the budgets of moderately volatile and trace elements, thereby influencing modeled mantle abundances. Carbon has long been proposed as a potential light element in Earth's core based on cosmochemical and experimental constraints, and more recently hydrogen has also been suggested from high-pressure experiments and seismological data \citep{Badro_2014,Li_2018,Li_2020,Hirose_2021,wang_2021}. However, the delivery of carbon and hydrogen during planetary accretion and its subsequent partitioning between the core and mantle remain highly uncertain. In addition, hydrogen incorporated into the mantle is susceptible to diffusive loss during ascent and partial melting, further obscuring its primary budget \citep{Xia_2002,xia_2013a,xia_2013b}. For these reasons, hydrogen and associated volatile-loss processes are not included in the present framework. Explicitly accounting for evaporation, escape, and volatile partitioning therefore represents an important direction for future model refinement \citep{johnston_1940,Sossi_2018,shaw_2023}.

Fourth, there is further consideration for metal-silicate differentiation. On the one hand, equilibration is parameterized using fixed equilibration fractions (Section~2.4.1 and Table~\ref{tab:equil_fraction}). However, nonideal mixing in the chemical system will inevitably impact element partitioning, a challenge that may require machine learning to address \citep{Zhang_2024}. On the other hand, equilibration efficiency likely varies with impact energy, melt fraction, and the dynamics of metal emulsification. Coupling the present framework with hydrodynamic or SPH impact simulations therefore represents an important avenue for improving the physical realism of late-stage differentiation \citep{rushmer_2005,Landeau_2021,Fischer_2025}.

\section{Conclusions}

In this work, We have presented a unified accretion-differentiation framework that links the dynamical growth of terrestrial planets with their chemical evolution, and apply it to study the formation of Earth and Mars simultaneously. By integrating N-body accretion simulations with impact-dependent metal-silicate equilibration, we reveal that the major differences between Earth and Mars analogs in mantle composition (e.g., Mg\#), FeO content, and core mass fraction emerge naturally from the three combined effects: 1) the provenance of accreted material (EF-EC-OC-CI mixing), 2) the bulk redox state of the planetary bodies, and 3) the pressure-temperature conditions during impact-driven equilibration.

Despite considerable model uncertainties and the stochastic nature of N-body dynamics, planetary compositions emerge as consequences of specific accretion and differentiation pathways. Recent studies have demonstrated that even small variations in accretion history or redox state can amplify into large differences in core-mantle composition \citep{fischer_2017,gu_2023,Fischer_2025}. 
Nevertheless, these differences remain robust for the Earth-Mars pair. Earth analogs are characterized by sustained accretion of reduced inner-disk material and frequent deep equilibration events, which enable efficient core growth while preserving a mantle with low FeO and high Mg\#. Mars analogs, in contrast, experience early outward scattering, accrete more oxidized material, and undergo shallower equilibration. This ultimately leads to higher mantle FeO, lower Mg\# and smaller cores. Critically, this distinct outcome for the two planet types requires no planet-specific ad hoc assumptions (e.g., distinct initial reservoirs or late-stage "special events") but instead reflects divergent evolutionary pathways inherent to the same protoplanetary disk.
 
This framework provides a physically grounded basis for interpreting the chemical diversity of terrestrial planets and is readily extended to rocky exoplanet populations. Future work will incorporate sulfur, volatile loss and more realistic equilibration physics, and extend the framework to alternative formation scenarios such as pebble accretion and migration, to assess the robustness of the inferred Earth-Mars divergence \citep{johansen_2021,halliday_2023,Johansen_2023a,Johansen_2023b,morbidelli_2024,Bizzarro2025}.

\begin{acknowledgements}

We thank Nicolas Dauphas, Shuai Kang, Menghua Zhu, Lu Pan for useful discussions. We also thank the anonymous referee for their valuable suggestions and comments.
B.L. is supported by the National Key R\&D Program of China (grant No. 2024YFA1611803), National Natural Science Foundation of China (grant Nos. 12222303, 12173035) and a startup grant of the Bairen program from Zhejiang University.  
The simulations in this work is supported by SilkRiver Supercomputer School of Physics and the Earth System Big Data Platform (School of Earth Sciences) at Zhejiang University.  
\end{acknowledgements}

\bibliographystyle{aa}
\bibliography{mybibliography}
\begin{appendix}
\nolinenumbers

\section{Bulk compositions of endmembers}
Table ~\ref{TabS1} summarizes the bulk compositions adopted for the meteoritic endmembers (EF, EC, OC and CI) and for the reference planetary bodies (Earth and Mars) used throughout this study. These compositions provide the initial chemical inventories for the accretion-differentiation calculations described in Sections 2 and 3. Mantle compositions are reported as major oxides, while core compositions are reported as elemental abundances. All values are normalized to MgO = 1 for consistency with Mg-normalized analysis used throughout the main text. Core elemental abundances are given as bulk contributions on the same normalization scale, rather than as fractions within the core.  
\begin{table}[h]

\centering
\caption{Bulk compositions of meteoritic end-members (EF, EC, OC, CI) and reference planetary bodies (Earth and Mars).}
\label{TabS1}
\setlength{\tabcolsep}{2pt}
\begin{tabular}{lcccccc}
\toprule
 & EF & EC & OC & CI & Earth & Mars \\
\midrule
\multicolumn{7}{l}{\textit{Silicate mantle composition (normalized to MgO = 1 by atomic)}} \\
\midrule
MgO    & 1.00000 & 1.00000 & 1.00000 & 1.00000 & 1.00000 & 1.00000 \\
Al$_2$O$_3$ & 0.05949 & 0.03337 & 0.03608 & 0.03966 & 0.04644 & 0.04578 \\
CaO    & 0.08686 & 0.04738 & 0.05352 & 0.05791 & 0.06732 & 0.06678 \\
FeO    & 0.00000 & 0.00000 & 0.49853 & 0.84387 & 0.11963 & 0.26604 \\
NiO    & 0.00046 & 0.00051 & 0.00036 & 0.04601 & 0.00357 & 0.00080 \\
SiO$_2$ & 0.40273 & 1.19006 & 1.06317 & 0.97042 & 0.79683 & 0.98460 \\
\midrule
\multicolumn{7}{l}{\textit{Core composition}} \\
\midrule
Fe (core) & 1.09704 & 0.89852 & 0.16618 & 0.00000 & 0.79309 & 0.43189 \\
Ni (core) & 0.04555 & 0.05062 & 0.03543 & 0.00000 & 0.04502 & 0.03826 \\
Si (core) & 0.16497 & 0.05347 & 0.00000 & 0.00000 & 0.07377 & 0.00000 \\
O (core)  & 0.00000 & 0.00000 & 0.00000 & 0.00000 & 0.06476 & 0.09861 \\
\bottomrule
\end{tabular}
\tablefoot{Mantle compositions are reported as major oxides, whereas core compositions are reported as elemental abundances (Fe, Ni, Si, O). All values are normalized to MgO = 1 (atomic) instead of being fully specified within complete planetary bodies; thus, a unique core-mantle fraction (CMF) is not defined before differentiation modeling. Core elemental abundances are expressed on the same normalized scale (i.e., as bulk contributions relative to MgO), rather than as fractions within the core.}
\end{table}
The hypothetical EF endmember represents a refractory-rich, highly reduced
inner-disk component that is not directly sampled by known meteorite
collections. Its defining features include elevated Al$_2$O$_3$ and CaO,
strong depletion in FeO, and reduced Si relative to EC and OC compositions.
Such a component is motivated by the preferential condensation and retention
of refractory lithophile elements at high temperatures in the innermost
protoplanetary disk, where more volatile silicates are expected to be partially
depleted.

Without invoking an EF-like component, the elevated Al and Ca abundances of
both Earth and Mars cannot be reproduced by mixtures of EC, OC and CI
materials alone, without simultaneously over-enriching volatile elements.
The EF composition therefore provides a physically motivated endmember that
allows the observed refractory element systematics of terrestrial planets to
be reconciled with their accretion histories.

\section{Mass-pressure scaling for core-mantle boundary conditions}
Table ~\ref{TabS2} lists the masses and corresponding core-mantle boundary (CMB) pressures of selected Solar System bodies spanning the terrestrial-planet mass range, including Vesta, the Moon, Mercury, Mars, Venus and Earth \citep{Lay_2008, Neumann_2014, Dumoulin_2017, steinbr_2018,brennan_2020,Tokle_2021}. These values provide empirical benchmarks linking planetary mass to interior pressure conditions relevant for metal-silicate equilibration.

Using the data in Table ~\ref{TabS2}, we derive a simple power-law fit between planetary mass and CMB pressure,
\begin{equation}
    P_{\mathrm{CMB}}(M) = 136.5 * M^{0.91}
\end{equation}
where $P_{\mathrm{CMB}}$ is expressed in GPa and $M$ is the planetary mass in units of Earth masses ($M_\oplus$). This relation captures the first-order increase of interior pressure with planetary mass across Solar System bodies.

In this study, the above scaling is used to estimate the CMB pressure of simulated planets when a fully self-consistent interior structure calculation is not performed. The resulting pressures are then employed to define metal-silicate equilibration conditions following the prescriptions described in Section 2 and 3. We emphasize that this fit is intended as an empirical interpolation across known Solar System bodies, rather than a detailed interior model, and is sufficient for capturing relative differences in equilibration depth among Earth analogs and Mars analogs. 

\begin{table}[h]
    \centering
    \caption{Masses (in Earth masses) and corresponding core--mantle boundary pressures
    of Solar System bodies.}
    \setlength{\tabcolsep}{2pt}
    \begin{tabular}{lcccccc}
    \hline
    & Vesta & Moon & Mercury & Mars & Venus & Earth  \\
    \hline
    Mass ($\rm M_{\oplus}$) & $4\times10^{-5}$ & 0.0123 & 0.055 & 0.107 & 0.815 & 1.0  \\  
    $P_{\mathrm{CMB}}$ (GPa) & 0.1 & 4.8 & 5.7 & 20.0 & 114 & 136 \\
    \hline
    \end{tabular}
    \label{TabS2}
\end{table}

\section{Selected Earth-Mars analog systems}
\subsection{Orbital architectures of selected systems}
\begin{figure}
\centering
\sidecaption
\includegraphics[width=\linewidth]{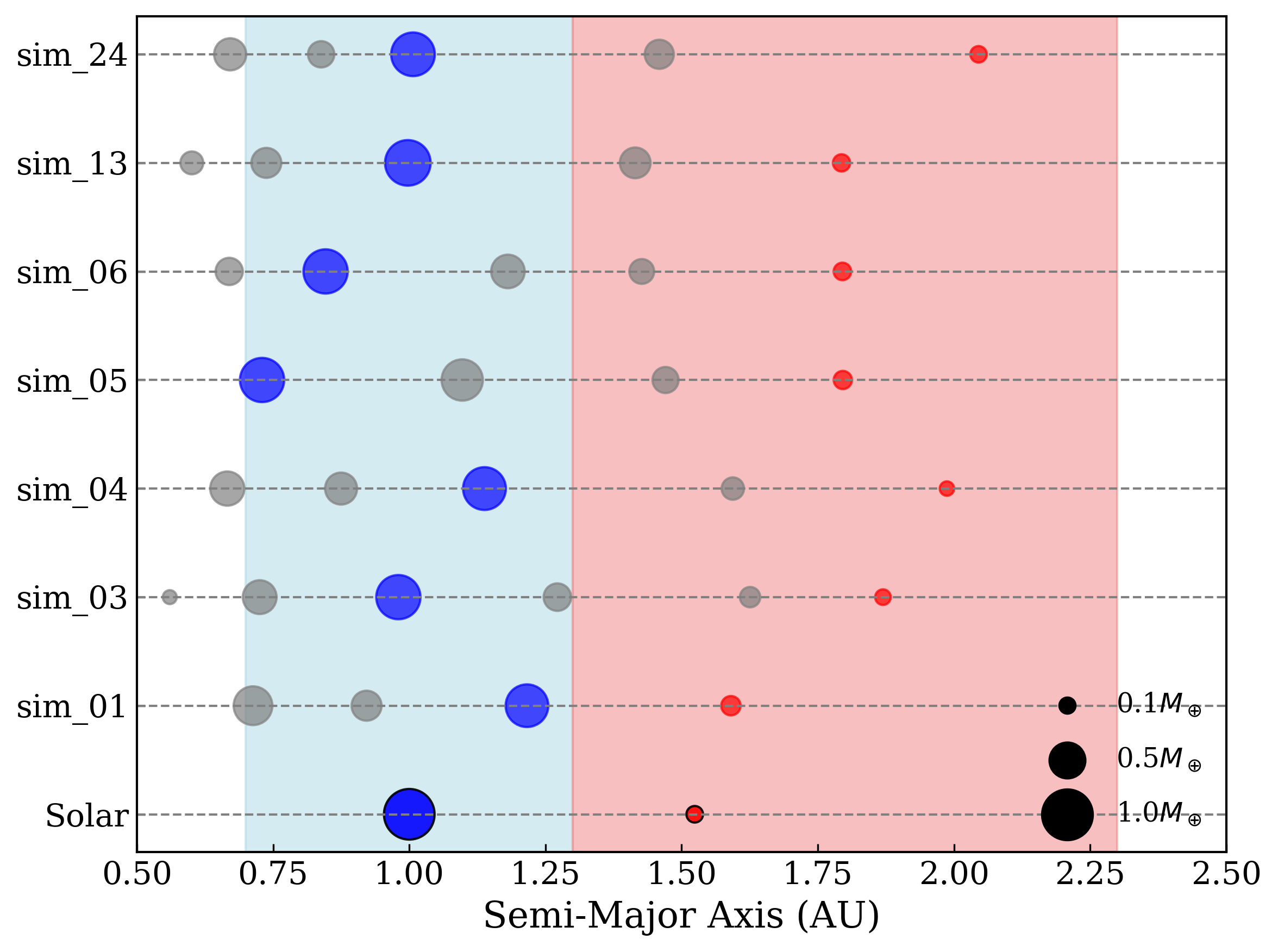}
\caption{Final orbital architectures of the seven selected Earth-Mars analog systems. Blue circles indicate Earth analogs and red circles indicate Mars analogs. Symbol size scales with planetary mass. The shaded regions indicate the adopted orbital selection ranges for Earth and Mars analogs. Only the Solar system analog reproduces a relatively clean Earth-Mars configuration without additional intermediate-mass planets between the two analogs, whereas the remaining systems retain one or more low-mass bodies between them.}\label{FigS1}
\end{figure}
FigS~\ref{FigS1} summarizes the final orbital architectures and compositional source fractions of the seven selected systems used in the Earth-Mars compositional analysis. All selected systems contain both an Earth analog and a Mars analog according to the criteria defined in Section 3. Among these systems, one reproduces a relatively clean Earth-Mars configuration without additional intermediate-mass bodies between the two analogs, whereas the remaining systems retain one or more low-mass planets between them. 
\subsection{Compositional source fractions}
Table ~\ref{TabS3} summarizes the accretion source fractions of the four compositional end-members (EF, EC, OC and CI) for all qualifying Earth analogs and Mars analogs produced in the simulations. Each row corresponds to an individual simulation system that satisfies the dynamical selection criteria described in Section 3.1, yielding one Earth analog near 1 AU and one Mars analog at larger heliocentric distance.

The source fractions represent the cumulative mass contribution of each end-member integrated over the full accretion history of the final planet. Earth analogs consistently accrete a dominant fraction of reduced inner-disk material, primarily EF and EC, with minor contributions from OC and negligible CI material. In contrast, Mars analogs systematically incorporate a larger proportion of oxidized material, particularly OC, and exhibit enhanced CI contribution relative to Earth analogs.

These differences in accretion source fractions quantify the bulk redox contrast between Earth analogs and Mars analogs that underpins the compositional trends discussed in Section 3 and 4. In particular, the higher EF-EC fractions of Earth analogs provide the reduced bulk composition required to offset FeO production associated with high-pressure metal-silicate equilibration, whereas the OC-CI-enriched inventories of Mars analogs favor higher mantle FeO, lower Mg\# and distinct core mass fractions.

\begin{table}[t]
\centering
\caption{Accretion source fractions of EF, EC, OC, and CI material for Earth analogs and Mars analogs in all simulation systems.}
\label{TabS3}
\setlength{\tabcolsep}{2pt}
\begin{tabular}{lcccc}
\toprule
Class & EF & EC & OC & CI \\
\midrule
Earth analogs(01) & 0.307455 & 0.322376 & 0.325313 & 0.044856 \\
Earth analogs(02) & 0.337147 & 0.316682 & 0.299985 & 0.046186 \\
Earth analogs(03) & 0.338333 & 0.318654 & 0.292929 & 0.050083 \\
Earth analogs(04) & 0.336662 & 0.327770 & 0.286813 & 0.048754 \\
Earth analogs(05) & 0.383062 & 0.289240 & 0.274682 & 0.053016 \\
Earth analogs(06) & 0.380293 & 0.304341 & 0.260900 & 0.054466 \\
Earth analogs(07) & 0.360849 & 0.313559 & 0.272917 & 0.052675 \\
Earth analogs(mean) & 0.349114 & 0.313232 & 0.287649 & 0.050005 \\ 
\midrule
Mars analogs(01)  & 0.193447 & 0.278059 & 0.381975 & 0.146519 \\
Mars analogs(02)  & 0.246682 & 0.197644 & 0.492446 & 0.063228 \\
Mars analogs(03)  & 0.267654 & 0.296220 & 0.315136 & 0.120990 \\
Mars analogs(04)  & 0.248530 & 0.219485 & 0.381364 & 0.150620 \\
Mars analogs(05)  & 0.323470 & 0.273924 & 0.316405 & 0.086202 \\
Mars analogs(06)  & 0.328984 & 0.241284 & 0.349453 & 0.080279 \\
Mars analogs(07)  & 0.209491 & 0.225301 & 0.442024 & 0.123185 \\
Mars analogs(mean)  & 0.259751 & 0.247417 & 0.382686 & 0.110146 \\
\bottomrule
\end{tabular}
\tablefoot{Fractions are computed from the cumulative accreted mass budget of each analog at $t=200$ Myr and sum to unity for each system.}
\end{table}

\section{Mean compositions expressed in weight percent}
In the main text, compositional evolution is primarily discussed using atomic proportions normalized to Mg, which facilitates direct comparison of accretion trends and redox-sensitive element paritioning. For comparsion with conventional geochemical datasets, Table~\ref{TabS4} additionally reports the mean mantle and core compositions of Earth analogs and Mars analogs in weight percent (wt\%). Values represent ensemble averages of all qualifying simulation systems, with uncertainties indicating the 1 $\pm \sigma$ dispersion among simulations. Reference compositions for Earth and Mars are also listed for comparison. Mantle compositions are expressed as oxide abundances, whereas core compositions are given as elemental abundances. The resulting core mass fractions (CMF) are also shown.
\longtab[3]{
\begin{table}[h]
\centering
\caption{Mean mantle and core compositions of Earth analogs and Mars analogs in weight percent (wt\%).}
\label{TabS4}
\setlength{\tabcolsep}{2pt}
\renewcommand{\arraystretch}{1.2}
\begin{tabular*}{0.9\textwidth}{
@{\extracolsep{\fill}}
l
S[table-format=2.2(3)]
S[table-format=2.2(3)]
S[table-format=2.2]
S[table-format=2.2]
}
\hline
{wt\%} & {Earth analogs} & {Mars analogs} & {Earth} & {Mars} \\
\hline
\multicolumn{5}{l}{\textit{Mantle}} \\
\hline
MgO        & 36.395        & 33.680        & 38.186 & 31.725 \\
Al$_2$O$_3$ & 4.051 \pm 0.064 & 3.592 \pm 0.100 & 4.485  & 3.674  \\
CaO        & 3.247 \pm 0.051 & 2.886 \pm 0.079 & 3.576  & 2.947  \\
FeO        & 9.183 \pm 0.535 & 12.116 \pm 2.711 & 8.142 & 15.044 \\
NiO        & 0.245 \pm 0.019 & 0.195 \pm 0.067 & 0.253 & 0.047 \\
SiO$_2$    & 46.878 \pm 1.233 & 47.530 \pm 1.508 & 45.358 & 46.564 \\
\hline
\multicolumn{5}{l}{\textit{Core}} \\
\hline
Fe         & 89.959 \pm 2.073 & 92.844 \pm 9.611 & 88.509 & 86.319 \\
Ni         & 5.095 \pm 0.054 & 5.988 \pm 0.201 & 5.280 & 8.035 \\
Si         & 3.918 \pm 0.497 & 0.954 \pm 0.453 & 4.141 & {--} \\
O          & 1.028 \pm 0.210 & 0.214 \pm 0.053 & 2.070 & 5.646 \\
\hline
CMF        & 0.302 \pm 0.008 & 0.250 \pm 0.006 & 0.320 & 0.250 \\
\hline
\end{tabular*}
\tablefoot{Values represent ensemble averages of all qualifying simulation systems, with uncertainties indicating the 1 $\pm \sigma$ dispersion among simulations. Reference compositions for Earth and Mars are also listed for comparison. Mantle compositions are reported as oxides, whereas core compositions are given as elemental abundances. The resulting CMF are also shown.}
\end{table}
}
\end{appendix}
\end{document}